\documentclass[9pt]{revtex4}
\usepackage{amsmath}
\usepackage{amsfonts}
\usepackage{amssymb}
\usepackage{mathrsfs}
\usepackage{graphicx, epstopdf}
\usepackage{multirow}
\usepackage{color}

\def\Q{\mathbf{Q}}
\def\P{\mathbf{P}}
\def\I{\mathbf{I}}
\def\z{\mathbf{z}}
\def\n{\mathbf{n}}
\def\m{\mathbf{m}}
\def\r{\mathbf{r}}
\def\x{\mathbf{x}}

\def\u{\mathbf{u}}

\begin{document}
\title{Solution landscapes of nematic liquid crystals confined on a hexagon}
\author{Yucen Han$^{1}$}
\author{Jianyuan Yin$^{2}$}
\author{Pingwen Zhang$^{2}$}
\author{Apala Majumdar$^{3}$}
\author{Lei Zhang$^{4}$}

\affiliation{$^1$Beijing International Center for Mathematical Research, Peking University, Beijing 100871, China.\\
$^2$School of Mathematical Sciences, Laboratory of Mathematics and Applied Mathematics, Peking University, Beijing 100871, China.\\
$^3$Department of Mathematics and Statistics, University of Strathclyde, Glasgow, G1 1XH, United Kingdom.\\
$^4$Beijing International Center for Mathematical Research, Center for Quantitative Biology, Peking University, Beijing 100871, China.}
\begin{abstract}
We investigate the solution landscape of a reduced Landau--de Gennes model for nematic liquid crystals on a two-dimensional hexagon at a fixed temperature, as a function of $\lambda$---the edge length.
This is a generic example for reduced approaches on regular polygons.
We apply the high-index optimization-based shrinking dimer method to systematically construct the solution landscape consisting of multiple solutions, with different defect configurations, and relationships between them.
We report a new stable T state with index-$0$ that has an interior $-1/2$ defect;
new classes of high-index saddle points with multiple interior defects referred to as H-class and TD-class saddle points;
changes in the Morse index of saddle points as $\lambda^2$ increases and novel pathways mediated by high-index saddle points that can control and steer dynamical pathways on the solution landscape.
The range of topological degrees, locations and multiplicity of defects offered by these saddle points can be used to navigate the complex solution landscapes of nematic liquid crystals and other related soft matter systems.
\end{abstract}
\maketitle

\section{Introduction} \label{sec:introduction}
Nematic liquid crystals (NLCs) are viscoelastic anisotropic materials that combine the fluidity of liquids with the long-range orientational order of solids.
The electro-optic properties of NLCs make them the working materials of choice for the multi-billion dollar liquid crystal display (LCD) industry.
NLCs have tremendous potential in nanoscience, biophysics and material design, all of which rely on modeling and computational methods for studying stable/unstable NLC states, switching mechanisms, dynamical processes on energy landscapes etc. \cite{de1993physics}.
These approaches are useful for soft matter systems in general, for the study of interfacial phenomena, active matter, polymers, etc. \cite{zhang2007morphology, han2020pathways,Teramoto2010Morphological, zhang2016recent}.

As mentioned above, NLCs are partially ordered materials, with locally preferred directions of averaged molecular alignment, referred to as \emph{nematic directors} \cite{de1993physics}. The existence of these distinguished directors leads to orientational order and directional physical, electro-optic and rheological properties. A particularly intriguing feature of experimentally observable NLC states in prototype geometries, are topological defects, which can be interpreted as discontinuities in nematic directors or localised regions of \emph{melting} or loss of orientational order. Defects have profound consequences for both static and dynamic phenomena in NLCs and yet several aspects remain poorly understood.
Defects commonly exist as isolated points or disclination lines in experiments, and are further classified by their topological degrees, such as the two-dimensional (2D) $\pm$1 and $\pm$1/2 point defects \cite{lavrentovich2012defects}. The topological degree is a measure of the defect strength i.e. the rotation of the leading nematic director around the defect core.
Defects can be unavoidable for NLCs in confinement i.e. confined to 2D and three-dimensional (3D) geometries with different types of boundary conditions, often due to the external fields \cite{oh1995electro}, geometrical and boundary constraints \cite{de2007point}, and sometimes energetic considerations etc. 
NLCs in confinement typically have multiple experimentally observable or stable states, and these stable states are often distinguished by distinct defect configurations. Recent years have seen a boom in mathematical studies for confined NLCs, particularly in a variational framework wherein we study stable observable NLC states as minimizers of an appropriately defined free energy.
The Landau-de Gennes (LdG) theory has been hugely successful in this respect, as shall be described in the next section and employed in this paper, see \cite{muvsevivc2006two, lubensky1998topological, onsager1949effects, bajc2016mesh} for studies on LdG solution landscapes, and \cite{majumdar2018remarks, majumdar2010landau, henao2017uniaxial, nguyen2013refined} for sophisticated studies of LdG energy minima which model stable experimentally observable states.

The confined NLC system can switch between different energy minima or stable states, by means of an external field, thermal fluctuations, and mechanical perturbations.
The switching requires the system to cross an energy barrier separating the two stable states, with an intermediate transition state.
The transition state is an index-$1$ saddle point, i.e., a stationary point of the energy functional such that the corresponding Hessian matrix has one and only one negative eigenvalue.
The transition state is the highest energy state along the transition pathway connecting the two stable states \cite{zhang2016recent}. There are typically multiple transition pathways, with distinct transition states, and the optimal transition pathway has the lowest energy barrier.
The reader is referred to \cite{kusumaatmaja2015free} for transition pathways on a square domain with tangent boundary conditions and to \cite{han2019transition} for transition pathways on a cylindrical domain with homeotropic/normal boundary conditions.

Transition states are the simplest kind of saddle points of the free energy. The Morse index of a stationary point of the free energy, is the number of negative eigenvalues of its Hessian matrix. In particular, energy minima or experimentally observable stable states are zero-index stationary points of the free energy with no unstable directions \cite{milnor1969morse}.  The analysis and numerical computation of high-index saddle points, with Morse index greater than $1$, is very challenging.
There are illustrative numerical results for multiple stationary points, in a 2D LdG framework, on a square domain in \cite{robinson2017molecular}. The authors apply the deflated continuation method to find 81 different stationary points of a reduced LdG energy, in a large 2D square domain, of which only six are stable (with zero Morse index) and the others are unstable saddle points.
However, the results in \cite{robinson2017molecular} are incomplete and the relationships between the different solutions are unclear.

Square wells are often used to study confined NLC systems, both experimentally and computationally, to elucidate the effects of geometry, boundary conditions, and material properties on LdG solution landscapes \cite{tsakonas2007multistable, kralj2014order, canevari2017order}.
In \cite{kralj2014order, canevari2017order}, the authors numerically discover and rigorously analyse the novel well order reconstruction solution (WORS) on a square domain, featured by a distinctive set of mutually orthogonal defect lines along the two square diagonals, with tangent boundary conditions which require the nematic directors to be tangent to the square edges. The WORS is globally stable, in a reduced LdG framework, for small nano-scale square domains; remains a stationary point of the free energy, for all square sizes, but loses stability as the square edge-length increases i.e. for larger square domains.
As the edge length increases, the diagonal defect lines become longer, and hence the LdG energy of WORS solution increases. The Morse index of the WORS increases as the edge length increases (in fact, we believe it is the highest index saddle point of the reduced LdG energy on square domains) and in \cite{yin2020construction}, the authors use the WORS as the parent state (the highest-index saddle point) and propose a general and efficient numerical method to construct the LdG solution landscape on a square domain i.e. a pathway map of connected solutions starting from a parent state, and connecting to admissible stable energy minima with zero Morse index, via intermediate saddle points and transition states.
This numerical study reveals several new saddle point solutions with multiple interior defects, which were previously unreported in the literature \cite{yin2020construction}.

The square domain is perhaps the most well studied amongst all regular 2D polygons, but it is special. For example, the WORS is not generic, with the two mutually orthogonal defect lines, for 2D polygons. In particular, in \cite{han2019reduced}, the authors show that the Ring solution, with a unique central point defect, is the generic stable solution for nano-scale regular polygons with $K$ edges, except for the square with $K=4$. Furthermore, for large regular $K$-polygonal domains, we have at least $[K/2]$ classes of stable states, distinguished by the locations of a pair of defects pinned at the polygon vertices. In contrast, for a disc (the limit of a polygon as $K \to \infty$), there is only one observable Planar Polar solution  featured by two interior nematic point defects along a disc diameter \cite{han2019transition}. In other words, the sharp vertices have a key role in stabilising multiple states, by means of stabilising different defect configurations. On these grounds, we choose the hexagon as a generic example of a 2D polygon with an even number of sides: the hexagon supports the generic Ring solution for small domains, does not support the special symmetric solutions exclusive to a square and is better suited to capture generic trends of the solution landscape, particularly with respect to geometrical parameters.  

In this paper, we apply a reduced LdG model to numerically compute the complex solution landscapes of NLCs inside a regular 2D hexagonal domain, with tangent boundary conditions.
The reduced LdG model, which captures the nematic orientational order in terms of a reduced LdG order parameter with two degrees of freedom, is suitable for 2D domains \cite{golovaty2017dimension}. The reduced LdG model effectively reduces to the Ginzburg--Landau model \cite{bethuel1994ginzburg} for superconductors at a fixed temperature, which is representative of temperatures below the NLC supercooling temperature.
In this reduced approach, we have one parameter---the domain size $\lambda^2$.
We largely vary $\lambda^2$ to construct a hierarchy of stationary points of the reduced LdG free energy on the hexagonal domain, including minimizers, transition states and high-index saddle points.
The hierarchy includes the previously reported Ring, P, and M solutions.
We find entirely new classes of high-index saddle points, e.g. the TD class (including index-$6$ saddle points) and the H class (including an index-$14$ saddle point), which cannot be identified on the square domain. Furthermore, unlike the WORS being the parent state on a square for all $\lambda^2$,
the parent state on a hexagon changes from the Ring solution, to the T135 saddle point, and to the index-$14$ H-type saddle point as $\lambda$ increases.
It is noteworthy that we find a new type of index-$0$ stable solution with an interior $-1/2$ defect, (the director rotates by $\pi$ radians around the defect core, and hence the topological degree of $-1/2$), referred to as the T solution.
We also observe certain numerical trends on how the director profile near the hexagon vertices (bend-like versus splay-like) affects the Morse index, as does the symmetry group of the saddle point.
A plethora of saddle point solutions gives us diverse possibilities for transition pathways and indeed, we illustrate the differences between transition pathways with transition states and transition pathways with high-index saddle points, e.g. two stable T solutions can be connected by an index-$8$ H-type saddle point. In some cases, transition pathways mediated by high-index saddle points can be more efficient for switching processes and this warrants further investigation.

The paper is organised as follows. In Section \ref{sec:framework}, we briefly review the reduced LdG framework for NLCs on 2D domains.
In Section \ref{sec:method}, we describe the numerical methods for computing index$-k$ saddle points and the algorithm for constructing the solution landscape.
In Section \ref{sec:results}, we systematically construct the solution landscapes with increasing complexity for $\lambda^2 = 70, 150$ and $600$ respectively, where the parameter $\lambda^2$ is a measure of the hexagonal domain size or the edge length.
In Section \ref{sec:comparison}, we compare the solution landscapes on square and hexagonal domains.
We finally present our discussion, conclusions and perspectives in Section \ref{sec:conclusion}.

\section{Landau--de Gennes theory}\label{sec:framework}

As a powerful continuum theory for NLCs, the LdG theory describes the NLC state with a macroscopic order parameter---the $\Q$-tensor, which is a symmetric, traceless $3\times 3$ matrix \cite{de1993physics}.
The NLC is said to be in the isotropic phase if $\Q = 0$, uniaxial if $\Q$ has a pair of degenerate nonzero eigenvalues, and biaxial if $\Q$ has three distinct eigenvalues.
A uniaxial $\Q$-tensor is often written compactly as
\begin{equation}\label{eq:uniaxial}
\Q = s\left(\n \otimes \n - \frac{\mathbf{I}}{3} \right),
\end{equation}
where $\n$ is the nematic director (i.e., the eigenvector with the non-degenerate eigenvalue) that models the single preferred direction of orientational ordering
and $\I$ is the identity matrix.

We work with a particularly simple form of the LdG energy
  \begin{equation}
       I[\Q]: = \int_{\Omega_\lambda} \left[\frac{L}{2}\left| \nabla \Q \right|^2 + f_b\left( \Q \right)\right]\mathrm{d}A, 
    \end{equation}
where $\Omega_\lambda$ is a 2D hexagonal domain with the edge length $\lambda$ 
and L is a positive elastic constant. We choose the simplest form of the LdG model by using the isotropic elastic energy for computational simplifications. $f_b$ is the bulk potential that drives the isotropic-nematic phase transition as a function of the temperature,
\begin{equation}\label{f_b}
    f_b = \frac{A}{2}\mathrm{tr} \Q^2 - \frac{B}{3} \mathrm{tr} \Q^3 + \frac{C}{4} (\mathrm{tr} \Q^2)^2,
\end{equation}
where $B, C$ are positive material-dependent constants, and $A=\alpha (T - T^*)$ is the rescaled temperature. 
For $A<0$, $f_b$ favours an ordered bulk uniaxial phase and $\mathcal{N}:=\{\Q\in \mathbb{M}^{3\times 3}:\Q=s_+(\n\otimes\n-\I/3)\}$ is the set of minimizers of $f_b$
 with
\begin{equation}\label{eq:splus}
s = s_+:= \dfrac{B + \sqrt{B^2 + 24 |A|C}}{4C}
\end{equation}
and $\n \in \mathbb{S}^2$.
We use MBBA as a representative NLC material and use fixed values $B = 0.64\times 10^4 N/m^2$ and $C=0.35\times10^4N/m^2$, which are reported in \cite{wojtowicz1975introduction}.

We nondimensionlize the system with $\bar{r} = r/\lambda$, and
the rescaled LdG energy functional is
\begin{equation}\label{eq:rescalede}
\bar{I}[\bar{\Q}]: = \int_{\Omega} \left[\dfrac{1}{2}\left| \bar{\nabla} \bar{\Q} \right|^2
 + \dfrac{\lambda^2}{L} f_b\left( \bar{\Q} \right) \right]\mathrm{d}\bar{A},
\end{equation}
where $\Omega$ is a regular polygon with the unit edge length.
In what follows, we drop the bars and all statements are in terms of the rescaled variables.
The corresponding Euler-Lagrange equations are:
\begin{equation}\label{eq:el}
\Delta \Q = \dfrac{\lambda^2}{L}\left(A \Q - B \left( \Q \Q - \dfrac{|\Q|^2}{3} \I \right) + C\left| \Q \right|^2 \Q\right).
\end{equation}
The physically relevant states are modelled as local or global energy minima subject to the imposed boundary conditions.

In \cite{golovaty2017dimension}, the authors prove that the physically relevant $\Q$-tensors on 2D domains have a fixed eigenvector $\z$, the unit vector in the $z$-direction, and can be written in terms of three variables $q_1, q_2, q_3$ as shown below:
\begin{equation}\label{eq:Qred}
\Q = q_1\left( \n\otimes \n - \m\otimes \m \right) + q_2 \left(\n\otimes \m + \m\otimes \n \right) + q_3 \left( 2\z\otimes \z - \n\otimes \n - \m\otimes \m \right)
\end{equation}
where $\n$ and $\m$ are orthonormal vectors in the $xy$-plane \cite{golovaty2017dimension}.
In other words, only three degrees of freedom out of five remain in a 2D framework.
Further, in \cite{canevari_majumdar_wang_harris}, the authors show that for $A = -\frac{B^2}{3C}$, $q_3$ is a constant for all physically relevant solutions of (\ref{eq:el}) of the form (\ref{eq:Qred}), subject to Dirichlet uniaxial tangent boundary conditions on the domain edges.
Hence, for $A=-\frac{B^2}{3C}$, we have a reduced description in terms of a reduced LdG tensor, $\P$, with only two degrees of freedom such that
\begin{equation}\label{eq:QP}
\Q =
\left(\begin{tabular}{cc|c}
\multicolumn{2}{c|}{\multirow{2}*{$\P\left(\r\right)+\dfrac{B}{6C}\I_2$}} & $0$ \\
\multicolumn{2}{c|}{} & $0$ \\ \hline
$0$ & $0$ & $-B/3C$ \\
\end{tabular}\right),
\end{equation}
where $\I_2$ is $2\times 2$ identity matrix.
In what follows, we track defects by using the nodal set or the zero set of $\P$ that is the set of uniaxial $\Q$-tensors with the negative order parameter about $\z$, as can be seen above, which is consistent with disorder in the plane of $\Omega$.

The corresponding reduced LdG energy is 
\begin{equation}\label{p_energy}
E[\P]: = \int_{\Omega} \left[\dfrac{1}{2}|\nabla \P|^2+ \dfrac{\lambda^2}{L} \left(-\dfrac{B^2}{4C}\mathrm{tr}\P^2+\dfrac{C}{4}\left(\mathrm{tr}\P^2\right)^2\right) \right]\mathrm{d} A.
\end{equation}
The Euler-Lagrange equations for $\P$ of \eqref{p_energy} are
\begin{equation}\label{euler_lagrange}
\left\{
\begin{aligned}
\Delta P_{11} &= \dfrac{2C\lambda^2}{L}\left(P_{11}^2+P_{12}^2-\frac{B^2}{4C^2}\right)P_{11},\\
\Delta P_{12} &= \dfrac{2C\lambda^2}{L}\left(P_{11}^2+P_{12}^2-\frac{B^2}{4C^2}\right)P_{12}.
\end{aligned}
\right.
\end{equation}
We further define a parameter
\begin{equation}\label{eq:lambdabar}
  {\bar{\lambda}}^2 = \dfrac{2C\lambda^2}{L}.
\end{equation}
We treat $C, L$ to be fixed material-dependent constants, and therefore $\bar{\lambda}^2$ is proportional to $\lambda^2$. For brevity, we drop the bar over $\lambda$ and use $\lambda^2$ to represent $\frac{2C\lambda^2}{L}$, as a measure of the domain size.
Such reduced descriptions have been hugely successful for 2D systems or thin three-dimensional (3D) systems, both for capturing the qualitative properties of physically relevant solutions and for probing into defect cores  \cite{brodin2010melting, bisht_epl,gupta2005texture,Mu2008Self,Igor2006Two}.
In a fully 3D system, there may be other classes of physically relevant solutions, such as escaped solutions, with additional degrees of freedom \cite{han2019transition, canevari_majumdar_wang_harris} and this will be pursued in further work.

\begin{figure}
    \begin{center}
         \includegraphics[width=0.3\columnwidth]{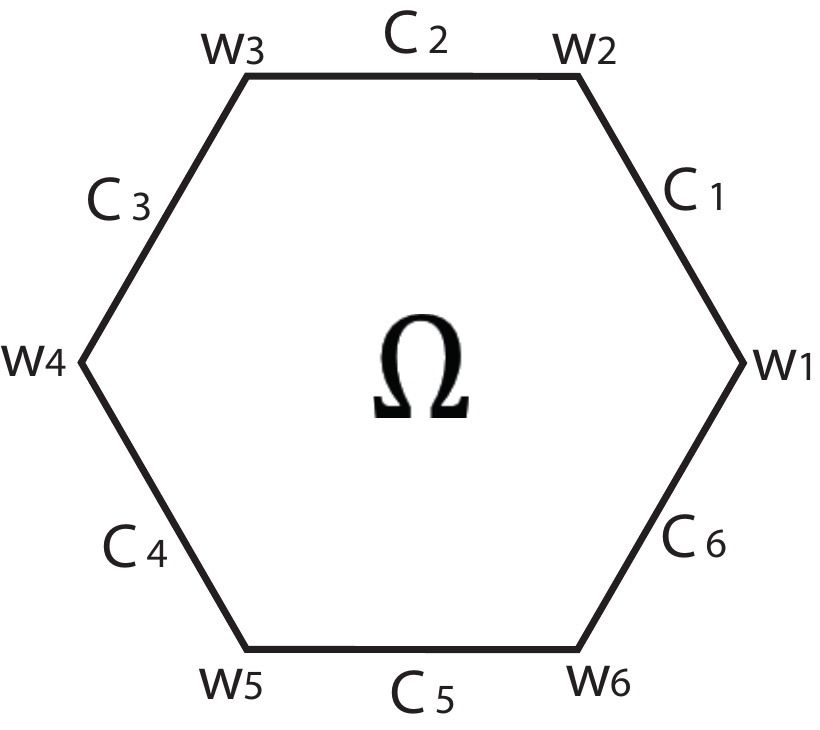}
         \caption{hexagonal domain}
         \label{figure:domain}
    \end{center}
\end{figure}

As illustrated in Figure \ref{figure:domain}, the regular hexagonal domain $\Omega$ is
centered at the origin with vertices
\begin{equation}\nonumber
w_k = \left(\cos \frac{\left(k-1\right)\pi}{3},\; \sin \frac{\left(k-1\right)\pi}{3}\right),\quad k = 1,\ldots,6.
\end{equation}
Starting from $\left(1,0\right)$, the edges are labeled counterclockwise as $C_1, \ldots, C_6$, and the distance between a point on $\partial\Omega$ and the vertices is defined as
\begin{equation}\nonumber
\mathrm{dist}\left(w\right) = \min\left\{\|w-w_k\|_2, k = 1,\ldots,6\right\}, \quad w \in \partial\Omega.
\end{equation}
The Dirichlet boundary conditions $\P = \P^b$ are imposed on the segments of edges as, away from the vertices
\begin{equation}\label{Pb}
\begin{aligned}
&P_{11}^b\left(w\right) = \alpha_k = -\dfrac{B}{2C}\cos\left(\dfrac{\left(2k-1\right)\pi}{3}\right), \\
&P_{12}^b\left(w\right) =  \beta_k = -\dfrac{B}{2C}\sin\left(\dfrac{\left(2k-1\right)\pi}{3}\right),
\end{aligned}
\quad w\in C_k, \mathrm{dist}\left(w\right)>\epsilon,
\end{equation}
where $0<\epsilon \ll 1/2$ is the size of the mismatch region.
We point out that the corresponding $\Q^b$ (associated with $\P^b$ in \eqref{eq:QP}) is in $\mathcal{N}$.
The value at the corner is the average of the two boundary conditions on the two intersecting edges.

The $\P$-tensor can also be expressed in terms of a scalar order parameter $s$ and an angle $\gamma$ as
\begin{equation}\label{P}
\P = s
\begin{pmatrix}\cos 2\gamma  & \sin 2\gamma  \\ \sin 2\gamma  &-\cos 2\gamma \end{pmatrix}
= 2s\left(\n\otimes\n-\dfrac{1}{2}\I_2\right),
\end{equation}
where $\n = \left(\cos\gamma,\sin\gamma\right)^\top$ is the nematic director in the plane, $s$ is a scalar order parameter that measures the degree of planar order about $\n$. The Dirichlet conditions \eqref{Pb} ensure that $\n$ is tangent to the edges, i.e., either parallel or antiparallel to the edges, so that this is a model 2D problem with tangent or planar boundary conditions.
In particular, one can use this representation to define the topological degree of $\P^b$ above, i.e., $\gamma$ changes by $2\pi$$N$ radians or $\n$ rotates by $2\pi$$N$ radians around $\partial \Omega$ for an integer or half integer $N$ and the corresponding topological degree of $\P^b$ is $N$.

\section{Numerical method}
\label{sec:method}
\subsection{HiOSD method}
It is a numerical challenge to find all stationary solutions, especially those saddle point solutions, of nonlinear partial differential equations such as the Euler-Lagrange equation in Eq. \eqref{euler_lagrange}.
In the past two decades, extensive numerical algorithms have been developed to compute saddle points, but most existing algorithms are designed to find index-1 saddle points. 
There are two popular approaches for searching index$-1$ saddle points. One is the path-finding methods, such as the nudged elastic band method \cite{jonsson1998nudged} and the string method \cite{weinan2002string}, and the other approach is the surface-walking methods, including the gentlest ascent dynamics \cite{weinan2011gentlest}, the dimer type method \cite{henkelman1999dimer,zhang2012shrinking,zhang2016optimization}, the eigenvector-following method \cite{doye2002saddle}, etc. 
Furthermore, many algorithms, e.g., the minimax method \cite{li2001minimax}, the deflation technique \cite{farrell2015deflation}, and the homotopy method \cite{mehta2011finding, hao2014bootstrapping} are dedicated to finding multiple stationary points of nonlinear equations, including both saddle points and minima, usually relying on an initial guess that deterministically leads to a stationary point.

In a recent work \cite{yin2019high}, the high-index optimization-based shrinking dimer (HiOSD) method is proposed to compute any-index saddle points, which can be viewed as a generalization of the optimization-based shrinking dimer method for searching index$-1$ saddle points \cite{zhang2016optimization}.
A minimax problem for an index$-k$ saddle point is formulated and then the maximal subspace is constructed by minimizing Rayleigh quotients simultaneously. Thus a dynamical system of the HiOSD is developed for finding an index$-k$ saddle point, and the stability analysis is performed to show that a linearly stable steady state of the HiOSD dynamical system is exactly an index$-k$ saddle point \cite{yin2019high}.

In what follows, we employ the HiOSD method to efficiently compute the stationary points (including both saddle points and minima) for the reduced LdG energy on a hexagon \eqref{p_energy}.
For a non-degenerate index-$k$ saddle point $\hat{\x}$, the Hessian $\mathbb{H}(\x)=\nabla^2 E(\x)$ at $\hat{\x}$ has exactly $k$ negative eigenvalues $\hat{\lambda}_1\leqslant \cdots \leqslant\hat{\lambda}_k$ with corresponding unit eigenvectors $\hat{\mathbf{v}}_1,\ldots , \hat{\mathbf{v}}_k$ satisfying $\big\langle\hat{\mathbf{v}}_j, \hat{\mathbf{v}}_i\big\rangle = \delta_{ij}$, $1\leqslant i, j \leqslant k$.
Define a $k$-dimensional subspace $\hat{\mathcal{V}}=\mathrm{span}\big\{\hat{\mathbf{v}}_1,\ldots, \hat{\mathbf{v}}_k\big\}$, then $\hat {\x}$ is a local maximum on a $k$-dimensional linear manifold $\hat{\x}+\hat{\mathcal{V}}$ and a local minimum on $\hat{\x}+\hat{\mathcal{V}}^\perp$, where $\hat{\mathcal{V}}^\perp$ is the orthogonal complement space of $\hat{\mathcal{V}}$.

The HiOSD dynamics for a $k$-saddle ($k$-HiOSD) is given as follows:
\begin{equation}\label{dynamics}
\left\{
\begin{aligned}
\beta^{-1}\dot{\x}   & =- \left(\I-2\sum_{j=1}^{k}\mathbf{v}_j \mathbf{v}_j^\top\right)\nabla E(\x), \\
\gamma^{-1}\dot{\mathbf{v}}_i & = -\left(\I-\mathbf{v}_i\mathbf{v}_i^\top-2\sum_{j=1}^{i-1}\mathbf{v}_j \mathbf{v}_j^\top\right)\mathbb{H}(\x)\mathbf{v}_i,\; i=1,\ldots,k,\\
\end{aligned}
\right.
 \end{equation}
where the state variable $\x$ and $k$ direction variables $\mathbf{v}_i$ are coupled, $\I$ is the identity operator and $\beta,\gamma>0$ are relaxation parameters.
The $k$-HiOSD dynamics \eqref{dynamics} is coupled with an initial condition:
\begin{equation}\label{othonormal}
\x(0)= \x^0\in\mathbb{R}^n,\quad \mathbf{v}_i(0)=\mathbf{v}_i^0\in\mathbb{R}^n,  i=1, \ldots, k,
\end{equation}
where $\mathbf{v}_1^0,\ldots,\mathbf{v}_k^0$ satisfy the orthonormal condition $\left\langle\mathbf{v}_i^0,\mathbf{v}_j^0\right\rangle = \delta_{ij}$, $i,j=1,2,\ldots,k$.
The first equation in \eqref{dynamics} describes a transformed gradient flow, which allows $\x$ to move along an ascent direction on the subspace $\hat{\mathcal{V}}$ and a descent direction on the subspace  $\hat{\mathcal{V}}^\perp$.
The second equation in \eqref{dynamics} is used to search for an orthonormal basis of $\hat{\mathcal{V}}$. Because the Hessian $\mathbb{H}(\x)$ is self-adjoint, we can simply take $\mathbf{v}_i$ as a unit eigenvector corresponding to the $i$th smallest eigenvalue of $\mathbb{H}(\x)$, which can be obtained from a constrained optimization problem,
\begin{equation}\label{minvi}
\min_{\mathbf{v}_i\in\mathbb{R}^n} \quad\langle \mathbb{H}(\x)\mathbf{v}_i, \mathbf{v}_i\rangle,\qquad
\mathrm{s.t.} \quad\langle \mathbf{v}_j,\mathbf{v}_i\rangle=\delta_{ij},\quad j=1,2,\ldots,i.
\end{equation}
Then we minimize the $k$ Rayleigh quotients \eqref{minvi} simultaneously by solving the second equation in \eqref{dynamics}.
To avoid direction-based calculation of Hessian, we use central difference schemes for directional derivatives to approximate Hessians by $k$ dimers centered at $\x$.
The $i$th dimer has a direction of $\mathbf{v}_i$ with a small dimer length $2l$ and $\mathbb{H}(\x)\mathbf{v}_i$ is approximated by
\begin{equation}\label{eqn:dimerappximation}
\mathbb{H}(\x)\mathbf{v}_i\approx \dfrac{\nabla E(\x+l \mathbf{v}_i) - \nabla E(\x-l\mathbf{v}_i)}{2 l}.
\end{equation}
HiOSD is a local-search algorithm for the computation of saddle point of arbitrary indices, driven by a given initial condition. The great advantage of HiOSD is that we can use it as an efficient tool for constructing the solution landscape, which gives a systematic approach for the search of saddle points and (local and global) minimizers, without random initial guesses. The connectivity of saddle points can be well established via the downward search and upward search methods, both of which are described in next subsection.

\subsection{Algorithm for constructing the solution landscape}
The solution landscape is a pathway map consisting of all stationary points and their connections.
Following the HiOSD dynamics, we construct the solution landscape by means of two algorithms:
a downward search that enables us to search for all connected lower-index saddles from an index-$m$ saddle;
an upward search with a selected direction to find the higher-index saddles, which drives the entire search to navigate up and down on the energy landscape \cite{yin2020searching}. 

{\it Downward search algorithm}: Given an index-$m$ saddle point $\hat{\x}$ and $m$ unit eigenvectors $\hat{\mathbf{v}}_1,\ldots,\hat{\mathbf{v}}_m$ corresponding to the $m$ negative eigenvalues $\hat{\lambda}_1\leqslant \ldots\leqslant\hat{\lambda}_m$ of the Hessian $\mathbb{H}(\hat{\x})$ respectively, we search for a lower index-$k$ ($k<m$) saddle point using HiOSD dynamics \eqref{dynamics}.
For the initial condition, we choose $\x(0) = \hat{\x}\pm\varepsilon \u$  for $\x$, where
we perturb the high-index saddle $\hat{\x}$ along the direction $\u$ with a small $\varepsilon$ to push the system away from the index-$m$ saddle $\hat{\x}$.
The direction $\u$ is a linear combination of $(m-k)$ vectors in the set of unstable directions $\{\hat{\mathbf{v}}_{k+1},\ldots,\hat{\mathbf{v}}_{m}\}$, whose negative eigenvalues have the smallest magnitudes.
The other $k$ eigenvectors $\hat{\mathbf{v}}_1,\ldots,\hat{\mathbf{v}}_k$ are the initial unstable directions $\mathbf{v}_i(0)$.
A typical choice of initial conditions in a downward search is $(\hat{\x}\pm\varepsilon\hat{\mathbf{v}}_{k+1}, \hat{\mathbf{v}}_1,\ldots,\hat{\mathbf{v}}_k)$.
Normally, a pair of index-$k$ saddles can be found, corresponding to the $\pm$ sign in the initial guess.
If the dynamics does not converge, 
a new initial condition is needed or another Morse index $k$ is attempted.

{\it Upward search algorithm}: We can also search for a higher index-$k$ saddle from an index-$m$ saddle $\hat{\x}$ ($m<k$) by using the HiOSD dynamics.
The index-$m$ saddle is $\hat{\x}$ with eigenvectors $\hat{\mathbf{v}}_1,\ldots,\hat{\mathbf{v}}_m$ corresponding to $m$ negative eigenvalues.
To search for a higher-index saddle, 
$(k-m)$ other unit eigenvectors $\hat{\mathbf{v}}_{m+1},\ldots,\hat{\mathbf{v}}_k$ corresponding to the smallest $k-m$ positive eigenvalues of the Hessian $\mathbb{H}(\hat{\x})$ are required.
The initial state $\x(0)$ is set as $\hat{\x}\pm\varepsilon\u$ where $\u$ is a linear combination of $\{\hat{\mathbf{v}}_{m+1},\ldots,\hat{\mathbf{v}}_k\}$, and a typical initial condition for $k$-HiOSD in an upward search is $(\hat{\x}\pm\varepsilon\hat{\mathbf{v}}_{k}, \hat{\mathbf{v}}_1,\ldots,\hat{\mathbf{v}}_k)$.

Each downward and upward search represents a pseudodynamics between a pair of saddle points, which presents valuable insights into transition pathways between stable and unstable solutions and the corresponding energy barriers.
By repeating the downward search or upward search, we are able to systematically find saddle points of various indices and uncover the connectivity of the complex solution landscape.
In \cite{han2019reduced}, the authors identify stable reduced equilibria in different regimes and use arc continuation methods to trace the corresponding solution branches. This method misses several solution branches, particularly disconnected branches and saddle point solutions.

In the next sections, we use the HiOSD method to study the solution landscape for different values of $\lambda^2$, and as such, discover new stable solutions, saddle point solutions, e.g. TD and H solutions with multiple interior defects.

\subsection{Spatial discretization on a hexagonal domain}
To maintain the symmetric properties of the hexagonal domain, we apply finite difference schemes over triangular elements to approximate the spatial derivatives, by analogy with the conventional discretization of a square domain \cite{fabero2001explicit}.
The hexagonal domain is divided into regular triangles with the edge length $h$. We choose the edge length of a regular triangle mesh to be $h=1/50$ for a fixed re-scaled regular hexagonal domain $\Omega$, centered at the origin with the first vertex pinned at $w_1=(1,0)$. We have tested the stability of the numerical results by refining the mesh size and the solutions are not sensitive to smaller choices of $h$.
The variable of interest $\phi$ is measured at the vertices.
The 2D Laplacian operator is
\begin{equation}
\Delta = \frac{\partial^2}{\partial x^2} + \frac{\partial^2}{\partial y^2} 
= \frac{2}{3}\left(\frac{\partial^2}{\partial \r_1^2} + \frac{\partial^2}{\partial \r_2^2} + \frac{\partial^2}{\partial \r_3^2}\right),
\end{equation}
where $\r_1 = (1,0), \r_2 = (\frac{1}{2},\frac{\sqrt{3}}{2}), \r_3 = (-\frac{1}{2},\frac{\sqrt{3}}{2})$, and the Laplacian can be approximated by
\begin{equation}
\Delta \phi(\x_0) \approx \dfrac{1}{h^2}\left(\dfrac{2}{3}\sum_{i=1}^3\left(\phi(\x_0+h\r_i)+\phi(\x_0-h\r_i)\right)-4\phi(\x_0)\right).
\end{equation}
The elastic potential $|\nabla \phi|^2$ in \eqref{p_energy} can also be approximated by
\begin{equation}
|\nabla \phi(\x_0)|^2\approx
\dfrac{1}{3h^2}\sum_{i=1}^3\left( (\phi(\x_0+h\r_i)-\phi(\x_0))^2+(\phi(\x_0-h\r_i)-\phi(\x_0))^2\right).
\end{equation}
We present our numerical results in the next section.

\section{Results}\label{sec:results}

\subsection{Typical solutions on the regular hexagon}
\begin{figure}
	\begin{center}
		\includegraphics[width=0.8\columnwidth]{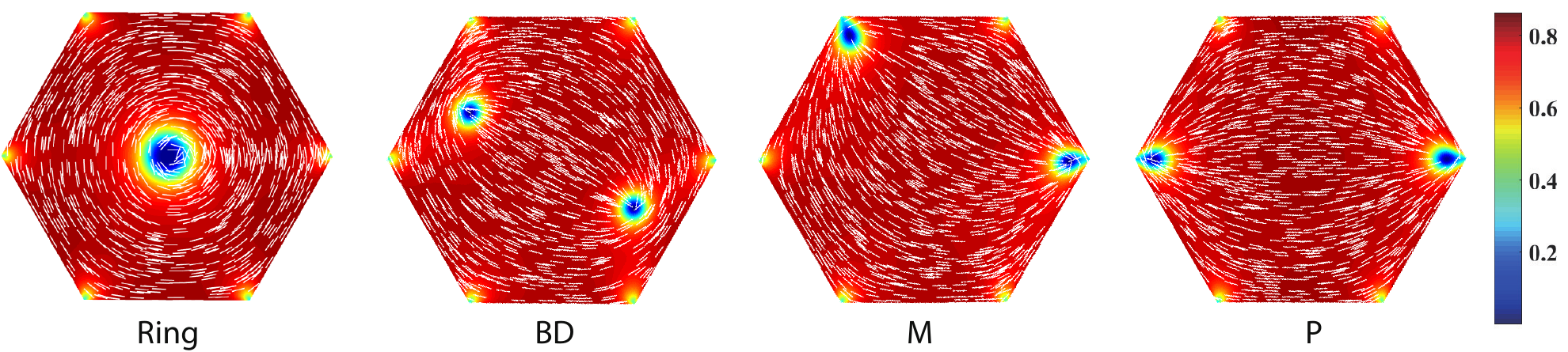}
		\caption{Four typical solutions: Ring, BD, M, and P at $\lambda^2 = 600$. The color encodes the 2D nematic order, $P_{11}^2+P_{12}^2$, and blue represents low nematic order manifested as defects. The white lines follow the planar nematic director. All subsequent figures have the same color bar for nematic order.}
		\label{figure:typical}
	\end{center}
\end{figure}

We plot some typical solutions of the reduced LdG model on a 2D hexagon in Figure \ref{figure:typical}, all of which were reported in \cite{han2019reduced}.
For small enough $\lambda$, we get the Ring solution with a $+1$ point defect at the centre, with a high degree of symmetry.
The Ring solution exists for all $\lambda$ and is globally stable in the reduced framework, in the $\lambda \to 0$ limit \cite{han2019reduced}.
The boundary distortion (BD) solution has two opposite +1/2 point defects near a pair of opposite edges, and this branch bifurcates from the Ring solution when the Ring solution loses stability, as $\lambda$ increases. 
There is an analogous bifurcation in a disc when the planar radial solution with a central $+1$ point defect bifurcates into planar polar solutions with two $+1/2$ point defects located along a diameter \cite{hu2016disclination}.
The M solutions have two point defects at vertices, which are separated by one vertex.
The P solutions are featured by a pair of diagonally opposite point defects and both the P and M solutions are stable for large enough $\lambda$.
There are no interior defects in the M and P solutions. 
The stable Ortho (O) solutions, with two adjacent point defects at adjacent vertices, only exist for very large values of $\lambda$ \cite{han2019reduced}, and they will not be studied in this paper.

\subsection{Solution landscape at $\lambda^2 = 70$}
\begin{figure}
\begin{center}
\includegraphics[width=0.7\columnwidth]{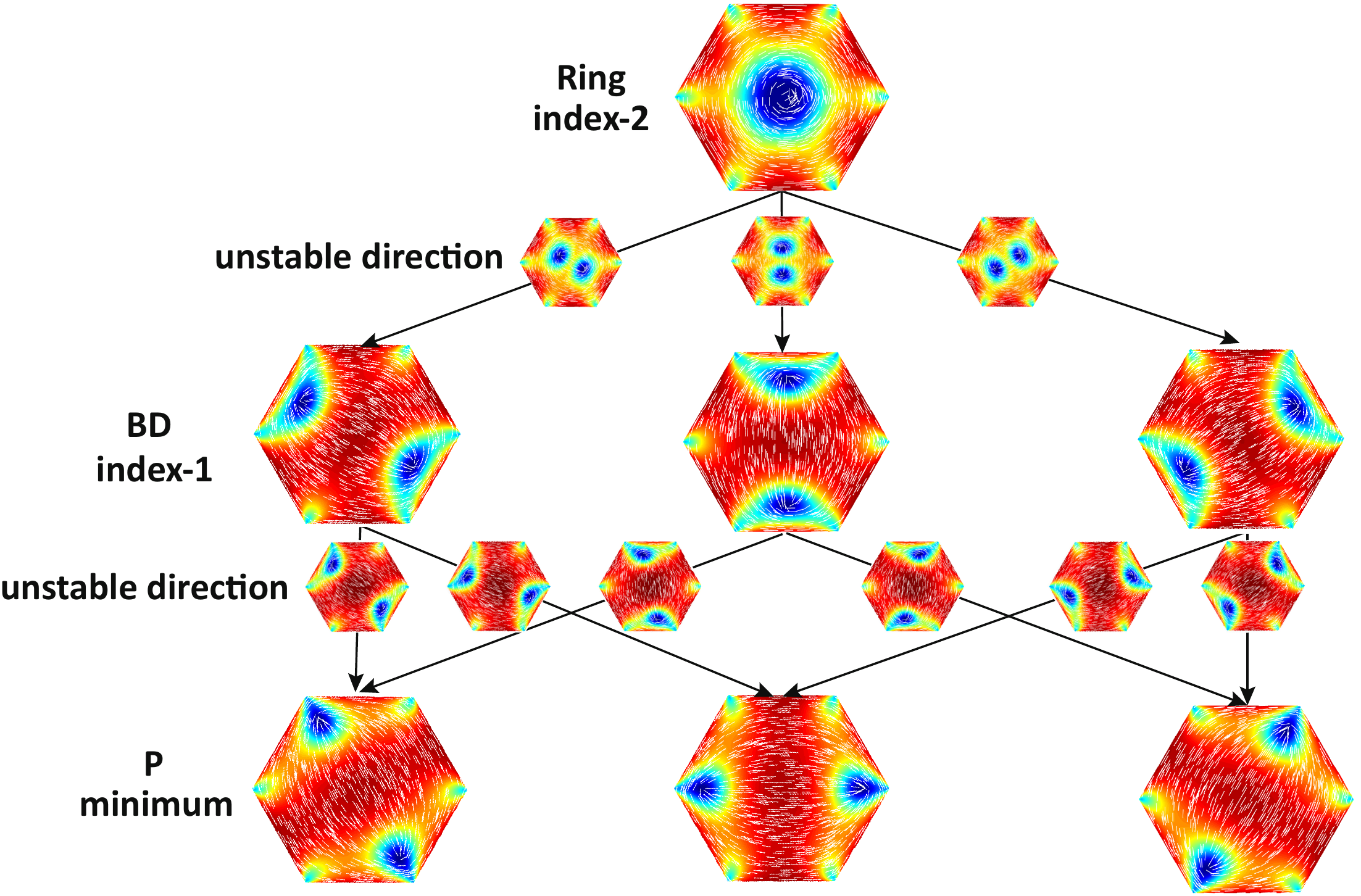}
\caption{Solution landscape at $\lambda^2=70$. 
Index-$2$ Ring is the parent state and connects to three index-$1$ BD solutions along its unstable directions. 
Each BD solution connects to two P minima along BD's single unstable direction.
}
\label{figure:70}
\end{center}
\end{figure}

When $\lambda^2$ is sufficiently small, the Ring solution is the unique stable solution \cite{han2019reduced}.
We perform an increasing $\lambda$ sweep for the Ring branch using Newton's method, the eigenvalues of the Ring solution monotonically decrease and we observe a pitchfork bifurcation at approximately $\lambda^2 \approx 10$, when we observe a transition from the unique stable Ring solution to multiple solutions \cite{troger2012nonlinear}. 
For $\lambda^2 \approx 10$, the Ring solution transitions from being a zero-index solution to a saddle point solution with index $2$ (with two equal negative eigenvalues), and we additionally have index-$1$ BD solutions and index-$0$ P solutions.
In general, we track bifurcations by tracking the indices of solutions; 
a change in the index is a signature of a bifurcation and a possible change of stability properties.
The solution landscape for $\lambda^2 =70$ is illustrated in Figure \ref{figure:70}, showing the relationships between Ring, BD, and P solutions.
The Ring solution is the parent state, i.e., the highest-index saddle point solution.
Following each unstable eigendirection of the Ring solution shown in Figure \ref{figure:70}, the central +1 point defect splits into two defects that relax around a pair of opposite edges, i.e., the BD solutions.
The two BD defects move from opposite edges to opposite vertices, following the unstable eigenvector of the BD solution and converging to the corresponding P solution.

\subsection{Solution landscape at $\lambda^2 = 150$}
\begin{figure}
\begin{center}
\includegraphics[width=0.7\columnwidth]{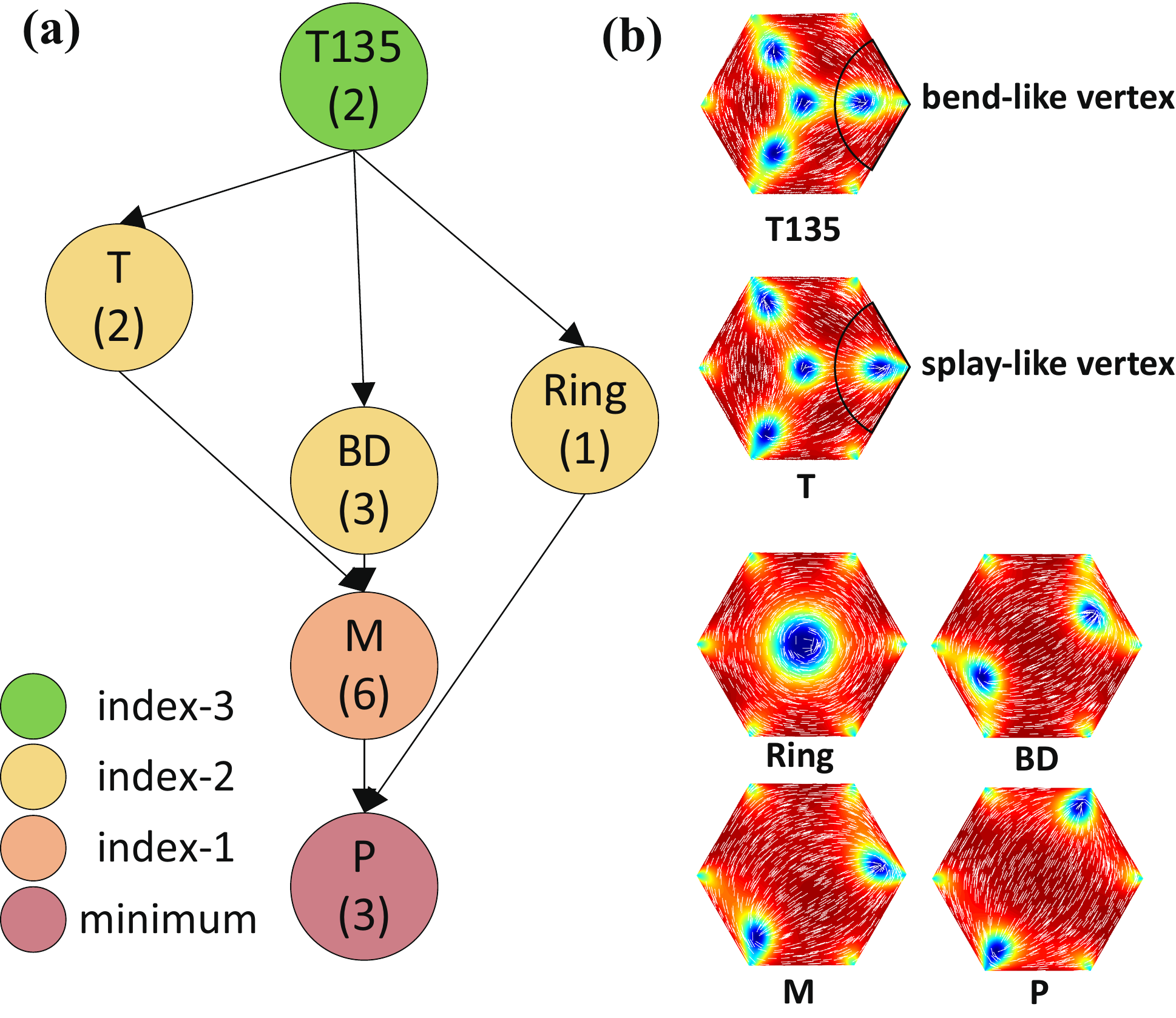}
\caption{(a) Solution landscape at $\lambda^2=150$. 
The colors of the nodes specify the Morse indices of saddle points.
The number in the parentheses indicates the number of solutions without taking symmetry into account. 
The height of a node approximately corresponds to its energy \eqref{dynamics}.
(b) The six configurations of solutions in (a). }
\label{figure:150}
\end{center}
\end{figure}

We obtain new saddle points in Figure \ref{figure:150} for $\lambda^2 = 150$.
The index-$3$ T135 replaces the index-$2$ Ring as a new parent state with two degenerate negative eigenvalues and one negative eigenvalue close to zero.
The new index-$2$ T solution has two degenerate negative eigenvalues and we deduce that these two saddle point solutions emerge from a saddle-node bifurcation, i.e., they appear suddenly without connecting with other branches in the bifurcation diagram.

Both T and T135 solutions have a central $-1/2$ point defect and three defects around alternate corners, with triangular symmetry. The configurations of T and T135 have subtle differences near the vertices. The defects of the T solution are closer to/pinned at the vertices of the hexagonal domain.
We can distinguish between the T and T135 solutions by examining the order parameter profiles near the vertices (\textit{bend-like vertex} or \textit{splay-like vertex} in Figure \ref{figure:150}(b)).
In the T solution, the +1/3 defect (around which the director $\n$, defined in \eqref{P}, rotates by $120^\circ$) is pinned to the vertex $w_1$ and $\n$ has a splay profile around $w_1$. 
We refer to such vertices with a pinned defect as a \textit{splay-like vertex}.
In T135, the +1/3 defect splits into a $+1/2$ interior defect and a $-1/6$ defect at $w_1$ (around which $\n$ exhibits a $60^\circ$ rotation), with a high order intermediate region between the interior $+1/2$ defect and $w_1$.
We refer to $w_1$ in T135 as a \textit{bend-like vertex}.
Regarding nomenclature, we use `T' for triangular symmetry and the indices $135$ to label the bend-like vertices in T135. 
In T135, the bend-like vertices are located at $w_1, w_3$ and $w_5$ respectively.
 We deduce that solutions with bend-like vertices have a higher Morse index than those with splay-like vertices.
The bend-like vertices have associated interior defects and we conjecture that this results in multiple unstable directions, and hence a higher Morse index.
The index-$2$ T solution has no bend-like vertices whereas the index-$3$ T135 solution has $3$ bend-like vertices. More examples are given for the H and TD solutions in the next subsection.

In comparison to $\lambda^2 = 70$, we also have new connections in Figure \ref{figure:150}(a) from T135 to T, BD and Ring respectively. 
The index-$1$ BD saddle solution bifurcates into an index-$2$ BD solution and an index-$1$ M saddle point. 
We have new connections between the BD, M and P solutions and the stable P solutions are connected by the index-$1$ M solutions in this case.
For $\lambda^2=150$, we obtain a total of $17$ solutions, without taking symmetry into account. If we take symmetry into account, there are only $6$ solutions and $7$ connections in Figure \ref{figure:150}. In terms of computational cost, by using Matlab 2018a on a Lenovo T450s laptop, the total CPU time needed for the construction of the complete solution landscape at $\lambda^2=150$ is $46$ minutes, and the average time for finding a new connection between two critical points is $6.5$ minutes.

\subsection{Solution landscape at $\lambda^2 = 600$}
\begin{figure}
\begin{center}
\includegraphics[width=0.9\columnwidth]{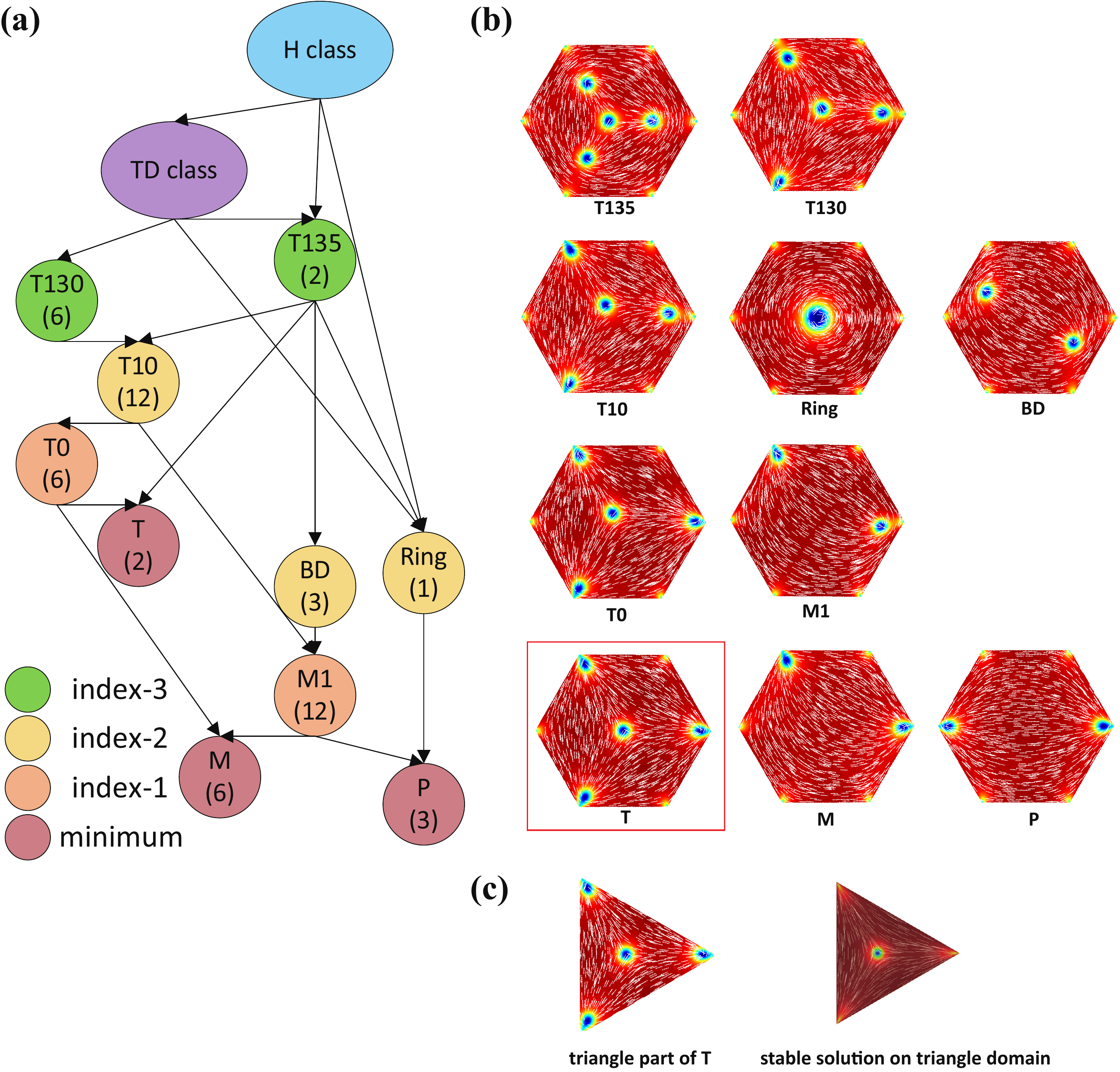}
\caption{(a) Solution landscape at $\lambda^2 = 600$. 
(b) The configurations corresponding to (a). 
(c) The triangle part of T solution on a hexagonal domain $\Omega$ and stable Ring solution on a triangle domain with $\lambda^2 = 450$.}
\label{figure:600}
\end{center}
\end{figure}

The solution landscape is quite complicated at $\lambda^2=600$ as shown in Figure \ref{figure:600}(a). 
There are three notable numerical findings in this regime: 
a new stable T solution with an interior $-1/2$ defect; 
new classes of saddle point solutions, H and TD, with high symmetry and high indices; 
new saddle points with asymmetric defect locations. 
We recover the T135 as an index-$3$ saddle point; 
the index-$2$ T solution bifurcates into an index-$0$ T and an index-$1$ T0;
we also observe the new index-$3$ T130 solution and an index-$2$ solution, labelled as T10.

Regarding nomenclature, the $0$ at the end of saddle point solutions T130, T10 and T0 indicates that the $-1/2$ defect is displaced from the centre and the other numbers label the bend-like vertices as before. 
We illustrate these configurations in Figure \ref{figure:600}(a-b), and the number in parentheses is simply the number of such configurations related to each other by symmetry.
The T135 and T solutions have three axes of reflection symmetry from the origin to $w_1$, $w_3$ and $w_5$ respectively; 
T130 is symmetric with respect to reflections about the line connecting the origin to $w_2$ and T0 is symmetric with respect to reflections about the line connecting the origin to the vertex $w_3$. 
The T10 saddle point has no axis of symmetry and there are $12$ distinct T10 solutions, which are related to each other by symmetry considerations.

The stable index-$0$ T solution is our first stable solution with an interior $-1/2$ defect at the centre of the hexagon, for $\lambda^2> 250$. 
The competing stable states, P and M, have defects pinned to vertices \cite{robinson2017molecular, han2019reduced}, and these vertex defects are a natural consequence of the tangent boundary conditions and topological considerations (the total topological degreeof the boundary condition is zero). 
We speculate that there may be other stable solutions with interior point defects, particularly on polygons with a greater number of sides, since the disc has stable planar polar solutions with two interior $+1/2$ defects. 
We also remark that the T solution on a hexagon (for large $\lambda$) is strongly reminiscent of the Ring solution on a regular triangle (Figure \ref{figure:600}(c)), as reported in \cite{han2019reduced}, which suggests that we can build new solutions by tessellating solutions on simpler building block-type polygons, such as the triangle and the square.

\begin{figure}
\begin{center}
\includegraphics[width=1\columnwidth]{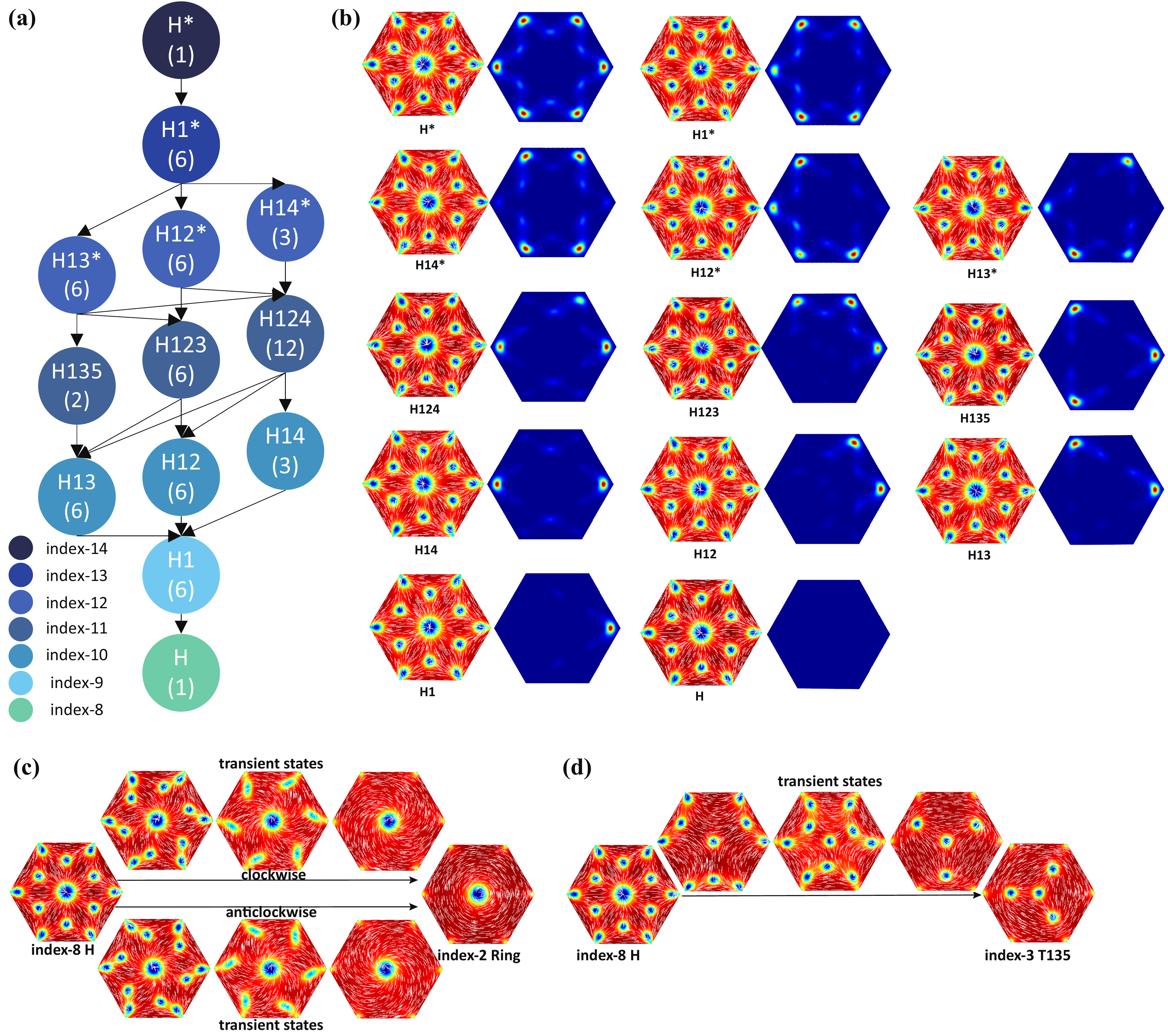}
\caption{(a) Solution landscape of the H class. 
(b) The corresponding configurations and plots of $|\P-\P^H|$, where $\P$ is any solution in the H class, and $\P^H$ is the index-$8$ H solution.
(c) From index-$8$ H solution to index-$2$ Ring solution following the HiOSD dynamics \eqref{dynamics}.
(d) From index-$8$ H solution to index-$3$ T135 solution following the HiOSD dynamics \eqref{dynamics}.}
\label{figure:H}
\end{center}
\end{figure}

\begin{figure}
\begin{center}
\includegraphics[width=\columnwidth]{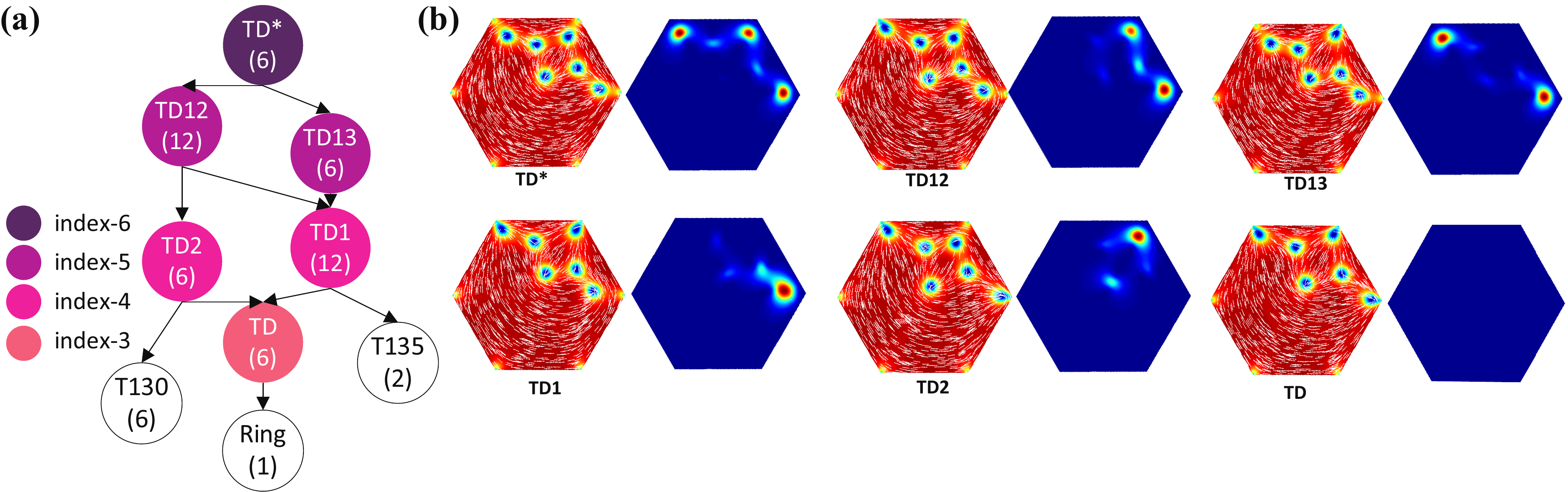}
\caption{(a) Solution landscape of the TD class.
 (b) The corresponding configurations of the TD class and plots of $|\P-\P^{TD}|$, where $\P$ is any solution in the TD class, and $\P^{TD}$ is the index-$3$ TD solution.}
\label{figure:TD}
\end{center}
\end{figure}

In the next paragraphs, we discuss saddle point solutions in the H class (Figure \ref{figure:H}) and TD class (Figure \ref{figure:TD}) that emerge from saddle-node bifurcations and pitchfork bifurcations, with high Morse indices and multiple interior defects.

{\it Solution landscape of the H class}:
We numerically find a new class of saddle point solutions, labelled as H-class solutions, which have Morse indices ranging from $8$ to $14$.
We plot the connectivity of these solutions in Figure \ref{figure:H}(a), and the corresponding configurations and their defect profiles in Figure \ref{figure:H}(b).

The parent state is the index-$14$ H* saddle point solution connecting to the lowest index-$8$ saddle point solution, labelled as H. 
Both of these states belong to the symmetry group $G_6:=\{S\in O(2):S\Omega \in \Omega\}$ (same as the Ring solution). Regarding nomenclature, we follow the same convention as before, i.e., H$135$ is an H-class saddle point with bend vertices at $w_1, w_3, w_5$ respectively.
The subscript $*$ labels the splay-like vertices (complement of bend-like vertices)
so that H* has no splay-like vertices whereas H has $6$ splay-like vertices. 
Other examples include the index-$13$ H1* with one splay-like vertex $w_1$, and the index-$9$ H1 solution has one bend-like vertex $w_1$; 
the index-$12$ H12* has two splay-like vertices $w_1$ and $w_2$, and the index-$10$ H12 solution has two bend-like vertices at $w_1$ and $w_2$; 
the index-$11$ H123 solution has three bend-like vertices pinned at $w_1$, $w_2$ and $w_3$ respectively.

The H-class saddle points look similar at first glance and we illustrate the subtle differences by plotting $|\P-\P^H|$, where $\P$ is a solution of Eq. \ref{euler_lagrange} in H-class and $\P^H$ is the index-$8$ H solution. 
The differences concentrate on the vertices with conspicuous red or white points in the dark blue background (Figure \ref{figure:H}(b)). 
These conspicuous points are localised near or at the bend-like vertices.
Numerically, we find that an index-$m$ solution in the H class has $(m - 8)$ bend-like vertices, e.g. the index-$8$ H solution has no bend-like vertices whereas the index-$14$ H* solution has $6$ bend-like vertices.

We follow the HiOSD dynamics in Figure \ref{figure:H}(c-d).
In Figure \ref{figure:H}(c), the unstable direction corresponding to the largest negative (closest to zero) eigenvalue of the H saddle point drives the corresponding director, $\n$ in (\ref{P}), to rotate clockwise or anticlockwise. 
Each $-1/2$ interior point defect merges with an adjacent splay-like vertex defect and disappears to give a defect-free profile around the vertex. 
The spiral transient state finally converges to the Ring solution.
In Figure \ref{figure:H}(d), the unstable direction corresponding to the fifth largest negative eigenvalue of index-$8$ H saddle point, drives three alternate $-1/2$ defects (in small triangles) and the $+1$ defect in the center of the hexagon to combine and annihilate, yielding a $-1/2$ defect.
The remaining alternate $-1/2$ defects are pushed to the edges, when the splay-like vertices disintegrate into bend-like vertices and from topological considerations, we are left with one central $-1/2$ defect and three symmetrically placed interior $+1/2$ defects. 
The final state is the T135 solution.

{\it Solution landscape of the TD class}:
We use TD as an abbreviation for ``triangle double" since TD solutions appear to be a superposition of two Ring solutions on a regular triangle, with two interior $-1/2$ point defects and an interior $+1/2$ point defect. 
The lowest-index saddle point solution in this class is the index-$3$ TD solution with no bend-like vertices. 
In general, a TD-type saddle point with $m$ bend-like vertices is index-$(m+3)$, so that the highest-index saddle point is TD* with $3$ bend-like vertices (the subscript $*$ has no vertex label attached to it which implies that there are no splay-like vertices). 
All saddle points in this class have three defective vertices, either bend-like or splay-like, as suggested by our numerical results and we illustrate the connectivity of this class in Figure \ref{figure:TD}(a). 
The TD class can also be connected to the T130, Ring and T135 saddle points as displayed in Figure \ref{figure:600}(a).

\subsection{Transition pathways between stable states}\label{sec:transition}

\begin{figure}
\begin{center}
\includegraphics[width=\columnwidth]{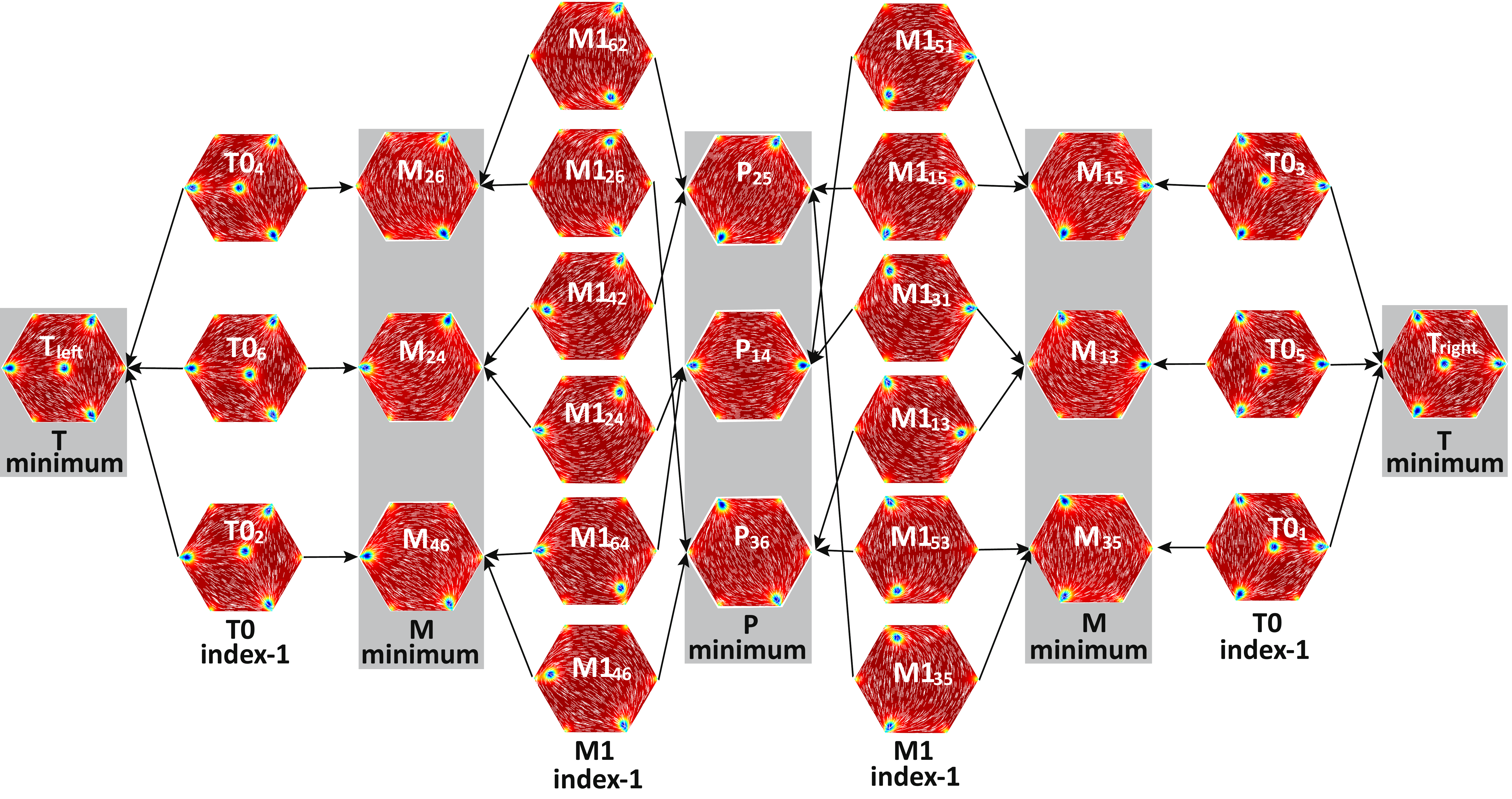}
\caption{The transition pathways between stable states including two T, six M and three P solutions at $\lambda^2 = 600$.}
\label{600_transition_pathway}
\end{center}
\end{figure}

We illustrate a comprehensive network of transition pathways between stable states including two T, six M and three P solutions at $\lambda^2 = 600$ in Figure \ref{600_transition_pathway}.
For clarity, we add subscripts to label the defect locations, since the hexagon is fixed and we need to cross energy barriers to switch between rotationally equivalent solutions (e.g. two different P or M solutions).
In Figure \ref{600_transition_pathway}, the T solution with defects pinned at $w_2, w_4, w_6$ is labelled as T$_{\mathrm{left}}$ and the T solution with defects pinned at $w_1, w_3, w_5$ is labelled as T$_{\mathrm{right}}$.
The subscripts in T0$_{i}$ simply identify the vertex closest to the displaced interior $-1/2$ defect. 
We use the label M1 to identify a class of index-$1$ saddle points, configurationally close to the M solutions, with one splay-like vertex and one bend-like vertex. 
We use two subscripts, i.e., M1$_{a,b}$ to identify the location of the bend-like and splay-like vertices respectively. 
In contrast, the stable M and P solutions have two splay-like vertices and we use subscripts, e.g. M$_{a,b}$, to locate the splay-like vertices.

Firstly, we remark that some stable and configurationally-close solutions are connected by a single transition state (index-$1$ saddle point) in Figure \ref{600_transition_pathway}.
For example, the transition state between T$_{\mathrm{left}}$ and M$_{26}$ is T0$_{4}$. 
The $-1/2$ center point defect in T$_{\mathrm{left}}$ moves towards the vertex $w_4$ and merges with the defect near $w_4$, as the configuration converges to M.
The transition state between M$_{26}$ and P$_{25}$ is M1$_{62}$. 
The defect at $w_6$ moves towards $w_5$ along the edge $C_5$ (between $w_5$ and $w_6$), and settles at $w_5$ yielding the stable P$_{25}$ state.

Secondly, two different M or P solutions cannot be connected by means of a single transition state, i.e., the transition pathway is typically composed of at least two transition pathways with an intermediate P or M state. 
From a practical perspective, this means that there is a high probability for the system to be trapped into the intermediate stable M or P state. 
For instance, one transition pathway between P$_{25}$ and P$_{36}$ is P$_{25}$--M1$_{35}$--M$_{35}$--M1$_{53}$--P$_{36}$.
The defect at $w_2$ moves towards $w_3$, yielding the stable M$_{35}$ and then the defect at $w_5$ moves towards $w_6$, yielding the stable P$_{36}$.
Similarly, one transition pathway between M$_{26}$ and M$_{35}$ is M$_{26}$--M1$_{26}$--P$_{36}$--M1$_{53}$--M$_{35}$, with an intermediate stable P$_{36}$ state. 
The transient dynamics involves the migration of the defect at $w_2$ towards $w_3$, which converges to P$_{36}$, followed by the motion of the defect at $w_6$ towards the vertex $w_5$ to yield the final stable state M$_{35}$.

The most complicated transition pathway appears to be the pathway between the two T solutions: T$_{\mathrm{left}}$ and T$_{\mathrm{right}}$.
Although T$_{\mathrm{left}}$ and T$_{\mathrm{right}}$ are two symmetric solutions related by a $60^{\circ}$ rotation, the switching process between T$_{\mathrm{left}}$ and T$_{\mathrm{right}}$ cannot be achieved by a simple rotation because the hexagonal domain is fixed.
In fact, one numerically computed transition pathway between T$_{\mathrm{left}}$ and T$_{\mathrm{right}}$ is T$_{\mathrm{left}}$--T0$_{4}$--M$_{26}$--M1$_{62}$--P$_{25}$--M1$_{15}$--M$_{15}$--T0$_{3}$--T$_{\mathrm{right}}$, where T0$_{4}$, M1$_{62}$, M1$_{15}$ and T0$_{3}$ are transition states (index-$1$ saddle points).
This shows that a transition between two energetically-close but configurationally-far T solutions may have to overcome four energy barriers and could be easily trapped by the stable M or P solutions.

In Figure \ref{600_transition_pathway}, each transition state can connect two stable minima by following its single unstable direction. 
This necessarily suggests that transition states cannot connect configurationally-far stable solutions, and hence multiple transition states are needed to connect configurationally-far stable states. 
This is computationally expensive and as suggested before, is not a reliable way of achieving switching because of the intermediate stable states. 
An alternative approach is to use higher-index saddle points with multiple unstable directions, to connect configurationally-far stable solutions. 
The multiple unstable directions give us greater control on the dynamical pathways and offer diverse possibilities, all of which give greater insights into the design and control of solution landscapes.

\begin{figure}
\begin{center}
\includegraphics[width=\columnwidth]{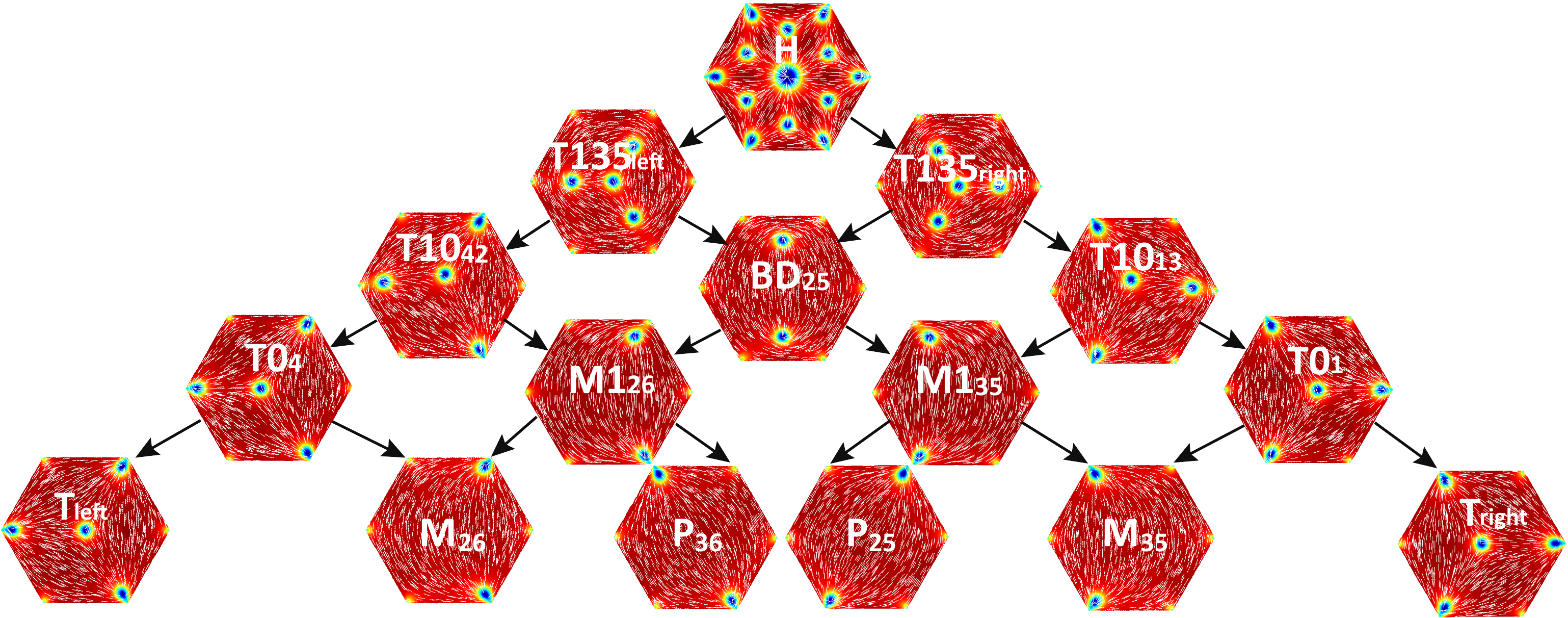}
\caption{Solution landscape starting from the H solution. 
All local minima such as T$_{\mathrm{left}}$, P$_{36}$, M$_{26}$, M$_{35}$, P$_{25}$, and T$_{\mathbf{right}}$ are connected by the index-$8$ H solution.}
\label{radar}
\end{center}
\end{figure}

Figure \ref{radar} shows how the different P, M, and T solutions are connected by high-index saddles.
For instance, the index-$1$ M1 solution connects the M and P solutions and the index-$1$ T0 solution connects the stable M and T solutions. 
However, two M solutions or two P solutions can be connected by the index-$2$ BD solution, e.g. M$_{26}$$\leftarrow$M1$_{62}$$\leftarrow$BD$_{25}$$\rightarrow$M1$_{53}$$\rightarrow$M$_{35}$ and P$_{25}$$\leftarrow$M1$_{62}$$\leftarrow$BD$_{25}$$\rightarrow$M1$_{53}$$\rightarrow$P$_{36}$.
The benefit of this pathway mediated by a high-index saddle point as opposed to a pathway with an intermediate stable state is that the system will not be trapped by the transient local minima along this pathway. 
Similarly, the M (or P) solutions and T solutions are connected by the index-$2$ T10 solution as follows: T$_{\mathrm{left}}$$\leftarrow$T0$_{4}$$\leftarrow$T10$_{42}$$\rightarrow$M1$_{26}$$\rightarrow$P$_{36}$ or M$_{26}$.
T$_{\mathrm{left}}$ and T$_{\mathrm{right}}$ solutions are configurationally far away from each other and are thus connected by an index-$8$ H solution: T$_{\mathrm{left}}$$\leftarrow$T135$_{\mathrm{left}}$$\leftarrow$H$\rightarrow$T135$_{\mathrm{right}}$$\rightarrow$T$_{\mathrm{right}}$.

Figure \ref{radar} shows that the index-$8$ H solution is the stationary point in the intersection of the smallest closures of two T, three P and six M solutions on the energy landscape.
The H solution is connected to every stable solution and we can thus construct dynamical pathways from the H solution to every individual stable solution.
Our numerical results highlight the differences between transition pathways mediated by index-$1$ saddle points and pathways mediated by high-index saddle points.
We deduce that index-$1$ saddle points are efficient for connecting configurationally-close stable solutions. 
For configurationally-far stable states, they are generally connected by multiple transition states and intermediate stable states, or in another way, connected by a high-index saddle point.

\section{Comparison with the solution landscape on a square}\label{sec:comparison}
\begin{figure}
\begin{center}
\includegraphics[width=0.8\columnwidth]{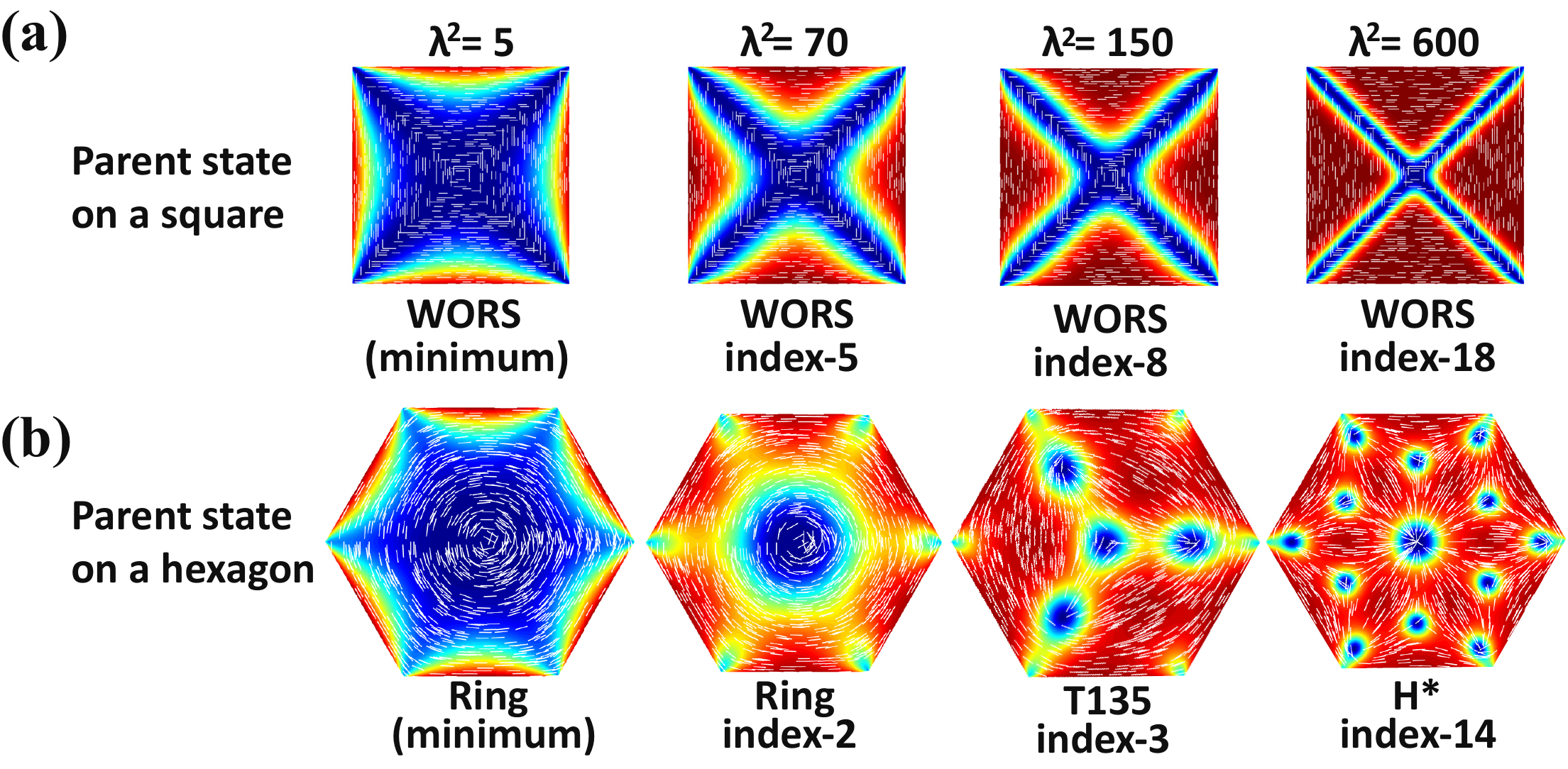}
\caption{Comparison of the parent states of the solution landscapes on the square (a) and the hexagon (b). 
The domain size $\lambda^2 = 5, 70, 150$, and $600$, respectively.}
\label{comparison}
\end{center}
\end{figure}

The reduced LdG solution landscape on a square domain with tangent boundary conditions has been studied in \cite{yin2020construction} and it is known that the WORS
 is the unique stable LdG equilibrium for small domain sizes, exists as a stationary point for all domain sizes, and is unstable for large domain sizes \cite{kralj2014order, canevari2017order}. 
The WORS is special in the sense that it is characterized by two isotropic defect lines along the square diagonals.
In Figure \ref{comparison} (a), the Morse index of the WORS increases with the domain size and the WORS is always the parent state for the solution landscapes on a square domain.
Intuitively, this is because the length of the diagonal defect lines increases as the domain size increases, and thus the WORS has an increasing number of unstable directions and an increasing Morse index, with the increasing square edge length.
The Ring solution, which is the analogue of the WORS on a hexagon, is index-$0$ for $\lambda$ small enough, and is an index-$2$ saddle point solution for larger $\lambda$, i.e., the Morse index does not increase with increasing $\lambda$.
The parent state with the highest index for the solution landscapes on a hexagon changes from the Ring solution to the index-$3$ T135 and index-$14$ H* when $\lambda^2 = 70, 150, 600$ (see Figure \ref{comparison} (b)) respectively , where T135 and H* solutions emerge through saddle-node bifurcations.
For the reduced LdG model on a hexagon, we have saddle-node bifurcations, stable solutions with interior point defects, and novel T, TD- or H-class states which cannot be found for any numerically tested value of $\lambda^2$ for the square domain.

Although the solution landscapes between a square domain and a hexagonal domain are quite different, there are some analogies.
The WORS and the Ring solution are unique stable solutions when $\lambda^2$ (the domain size) is small enough on a square and hexagon, respectively.
The BD on a square or hexagon is the first unstable solution which bifurcates from the parent state, the WORS or the Ring solution respectively.
The D and R solutions are stable solutions on a square \cite{robinson2017molecular} analogous to P and M solutions on a hexagon, when $\lambda^2$ is large enough \cite{han2019reduced}.
For the solution landscape on a square, WORS$\rightarrow$BD$\rightarrow$D is analogous to Ring$\rightarrow$BD$\rightarrow$P connections on a hexagon at $\lambda^2=70$ in Figure \ref{figure:70}.
We believe that the hexagon is a more generic example of a regular polygon with an even number of sides than a square and hence, we expect the qualitative aspects of our numerical study to extend to other regular polygons with an even number of sides.

\section{Discussion and conclusion}\label{sec:conclusion}

We investigate the solution landscape of a reduced LdG model on a regular 2D hexagon with tangent boundary conditions, as a prototype problem concerning nematic equilibria on regular 2D domains. 
We study reduced LdG energy minima in \cite{han2019reduced}, and report the Ring solution for nano-scale hexagons, and the P and M stable states for larger micron-scale hexagons, including a bifurcation diagram for the solution branches as a function of $\lambda^2$, i.e., we trace the continuation of these solution branches as a function of the hexagon size.
We go much further in this manuscript in the sense that we focus on the solution landscape including both minima and saddle points, and their relationships.
We first present illuminating numerical results on how the Morse index of solutions changes with $\lambda$, i.e., the Ring solution with a stable central $+1$ defect is index-$0$ for small $\lambda$ and is index-$2$ for large $\lambda$. 
Similarly, the BD solution branch changes from index-$1$ to index-$2$ as $\lambda$ increases. 
This is an interesting numerical example on how a given solution becomes more unstable as a function of the geometry. 
We observe new solution branches, e.g. TD- and H-class saddle points through saddle-node bifurcations, and quite importantly, we report two new stable T solutions with a central $-1/2$ defect, surrounded by three splay-like vertices. 
In fact, the T solution gains stability as $\lambda$ increases, with  a transition from index-$2$ to index-$0$ as $\lambda$ increases, and is reminiscent of the Ring solution for a regular triangle superimposed on a hexagonal domain \cite{han2019reduced}. 
This raises the pertinent question---can we construct reduced solutions on complex geometries by using reduced solutions on simpler geometries as a building block? 
We illustrate the solution landscapes at three representative values of $\lambda^2$ and strongly speculate that there are more stable solutions, symmetric and asymmetric saddle point solutions, tessellated solutions as $\lambda$ increases. 
Further, since the Dirichlet boundary condition is either topologically trivial or has a unit degree, we could build a further hierarchy of exotic reduced solutions by exploiting the topological degree of the tangent boundary condition.

A further innovative aspect of our study are insightful numerical results on dynamical pathways. 
We present several informative examples of transition pathways with a single transition state, multiple transition states and pathways mediated by higher-index saddle points. 
We believe that pathways mediated by high-index saddle points give greater possibilities for transition pathways, without the risk of being trapped into metastable states, and we can control the dynamical pathways by manipulating different unstable directions. 
The selective mechanism for dynamical pathways, in other words, how a system chooses between multiple dynamical pathways connecting a pair of equilibria, remains an open problem of practical significance for liquid crystal devices.

There are deep analytic issues associated with saddle points of nonlinear and non-convex functionals, such as the reduced LdG free energy.
We speculate that the saddle points of the reduced LdG model can be interpreted as minimizers of constrained problems or appropriately defined Dirichlet problems, as the WORS in \cite{canevari2017order}.
Advances in this direction would lead to new control strategies for confined NLC systems, since we could propose conditions that would stabilise the saddle point solutions, some of which have multiple interior and boundary defects.
Secondly, the hexagon is a generic regular polygon with an even number of sides.
Our numerical results, at least from qualitative aspects, will carry over to arbitrary regular polygons with an even number of sides, and indeed to a large number of phase field models on 2D polygons with planar boundary conditions.
Our results are indeed a consequence of the symmetry of the domain and the mathematical formulation, and as such, reveal certain universal trends of complex solution landscapes with regards to geometry, confinement and defects.

For a regular hexagonal domain, the critical points obtained are isolated, i.e., the corresponding Hessian has no zero eigenvalues. However, if the domain is a disc \cite{han2019transition}, 
the solution like the Planar Polar solution, has rotational invariance, hence its Hessian has one zero eigenvalue, so that it is a degenerate critical point (i.e., not an isolated critical point). The degeneracy of a critical point will certainly affect its numerical computation. In practice, the HiOSD method can compute such degenerate critical points by including the zero eigenvectors into the unstable directions. For example, a $k$-saddle with $m$ zero eigenvalues can be found using $(k+m)$-HiOSD. In a very recent work \cite{yin2020transition}, we apply this approach to search for critical points that are not isolated and the HiOSD method performs very well for identifying the degenerate critical points.
Furthermore, because of the hierarchical structure of the solution landscape, each HiOSD is independent of the others and the downward or upward search algorithms can naturally be parallelized, which can reduce the computational time greatly.

We also remark that our results, though restricted to a 2D setting, will exist in a 3D setting too, for example on a well with a 2D hexagon as cross-section \cite{canevari_majumdar_wang_harris}. 
In other words, these critical points exist as translationally invariant 3D critical points e.g. on 3D domains with free boundary conditions on the top and bottom surfaces, but they may not be energy minimizing in a 3D framework or the index of a 2D saddle point may be different in the 3D setting.
In fact, the Ring solution will exist as a translationally invariant solution on a 3D well, for arbitrary well heights, and will be globally stable for $\lambda$ small enough, independently of the well height. 
It is not clear if the Morse index of the saddle points will increase in a 3D setting and we expect this to be strongly dependent on the boundary conditions in a 3D setting. 
A reduced LdG tensor only has two degrees of freedom whereas the full LdG tensor in 3D has five degrees of freedom, allowing for more instabilities and on these grounds, we speculate that the T solution on a hexagon may not be stable in a fully 3D framework.

The results in this paper pose several challenging analytic and numerical questions. 
Can we obtain bounds for the Morse index as a function of $\lambda$? 
Is there an upper bound independent of $\lambda$, which would also impose restrictions on the complexity of the solution landscape with increasing $\lambda$? 
How can we check if the constructed solution landscape is complete? Is it possible that there exist isolated saddle points that are unconnected with the current solution landscape? Such questions warrant comprehensive theoretical and numerical studies in the future.
Other natural generalizations include the effects of the asymmetry in geometry, elastic anisotropy, and the boundary conditions. Preliminary work shows that elastic anisotropy perturbs the symmetry of the problem so that we may lose the Ring solution and the WORS with elastic anisotropy. 
In \cite{luo2012multistability}, the authors consider both strong and weak anchoring for planar liquid crystal wells with square cross-sections. Certain stable solutions (e.g. the rotated solutions) only exist for anchoring strengths larger than a critical anchoring strength. This suggests that the boundary conditions affect the solution landscapes, both in terms of the stable states and the saddle points.
We defer these investigations of the dimensionality, geometrical features and material anisotropy to future work.

{\bf Acknowledgment}
This work was supported by National Natural Science Foundation of China No. 11861130351, 21790340, 11421101, and the Royal Society Newton Advanced Fellowship awarded to Lei Zhang and Apala Majumdar. J. Y. acknowledges the support from the Elite Program of Computational and Applied Mathematics for Ph.D. Candidates of Peking University.


{\bf References}

\begin{thebibliography}{10}
\bibliographystyle{apalike}

	\bibitem{de1993physics}
	P.~G.~de Gennes and J.~Prost.
	\newblock {\em The physics of liquid crystals}, volume~83.
	\newblock Oxford university press, 1995.
	
	\bibitem{zhang2007morphology}
	L.~Zhang, L.~Q. Chen, and Q.~Du.
	\newblock Morphology of critical nuclei in solid-state phase transformations.
	\newblock {\em Physical Review Letters}, 98(26):265703, 2007.
	
	\bibitem{han2020pathways}
	Y.~C. Han, Z.~R. Xu, A.~C. Shi, and L.~Zhang.
	\newblock Pathways connecting two opposed bilayers with a fusion pore: a
	molecularly-informed phase field approach.
	\newblock {\em Soft Matter}, 2020.
	
	\bibitem{Teramoto2010Morphological}
	T.~Takashi and N.~Yasumasa.
	\newblock Morphological characterization of the diblock copolymer problem with
	topological computation.
	\newblock {\em Japan Journal of Industrial $\&$ Applied Mathematics},
	27(2):175--190, 2010.
	
	\bibitem{zhang2016recent}
	L.~Zhang, W.~Q. Ren, A.~Samanta, and Q.~Du.
	\newblock Recent developments in computational modelling of nucleation in phase
	transformations.
	\newblock {\em NPJ Computational Materials}, 2:16003, 2016.
	
	\bibitem{lavrentovich2012defects}
	O.~D. Lavrentovich, P.~Pasini, C.~Zannoni, and S.~Zumer.
	\newblock {\em Defects in liquid crystals: Computer simulations, theory and
		experiments}, volume~43.
	\newblock Springer Science \& Business Media, 2012.
	
	\bibitem{oh1995electro}
	M.~Oh-e and K.~Kondo
	\newblock Electro-optical characteristics and switching behavior of the in-plane switching mode.
	\newblock {\em Applied physics letters}, 67(26):3895--3897, 1995.

	\bibitem{de2007point}
	G.~de Luca, G. and A. D.~Rey
	\newblock Point and ring defects in nematics under capillary confinement
	\newblock {\em The Journal of chemical physics}, 127(10):104902, 2007.
	
	\bibitem{lubensky1998topological}
	T.~C. Lubensky, D.~Pettey, N.~Currier, and H.~Stark.
	\newblock Topological defects and interactions in nematic emulsions.
	\newblock {\em Physical Review E}, 57(1):610, 1998.
	
	\bibitem{onsager1949effects}
	L.~Onsager.
	\newblock The effects of shape on the interaction of colloidal particles.
	\newblock {\em Annals of the New York Academy of Sciences}, 51(1):627--659,
	1949.
	
	\bibitem{bajc2016mesh}
	I.~Bajc, F.~Hecht, and S.~{\v{Z}}umer.
	\newblock A mesh adaptivity scheme on the {Landau--de Gennes} functional
	minimization case in {3D}, and its driving efficiency.
	\newblock {\em Journal of Computational Physics}, 321:981--996, 2016.
	
	\bibitem{majumdar2018remarks}
	A.~Majumdar and Y.~W. Wang.
	\newblock Remarks on uniaxial solutions in the {Landau--de Gennes} theory.
	\newblock {\em Journal of Mathematical Analysis and Applications},
	464(1):328--353, 2018.
	
	\bibitem{majumdar2010landau}
	A.~Majumdar and A.~Zarnescu.
	\newblock {Landau-de Gennes} theory of nematic liquid crystals: the
	{Oseen-Frank} limit and beyond.
	\newblock {\em Archive for Rational Mechanics and Analysis}, 196(1):227--280,
	2010.
	
	\bibitem{henao2017uniaxial}
	D.~Henao, A.~Majumdar, and A.~Pisante.
	\newblock Uniaxial versus biaxial character of nematic equilibria in three
	dimensions.
	\newblock {\em Calculus of Variations and Partial Differential Equations},
	56(2):55, 2017.
	
	\bibitem{nguyen2013refined}
	L.~Nguyen and A.~Zarnescu.
	\newblock Refined approximation for minimizers of a {Landau-de Gennes} energy
	functional.
	\newblock {\em Calculus of Variations and Partial Differential Equations},
	47(1-2):383--432, 2013.
	
	
	\bibitem{kusumaatmaja2015free}
	H.~Kusumaatmaja and A.~Majumdar.
	\newblock Free energy pathways of a multistable liquid crystal device.
	\newblock {\em Soft matter}, 11(24):4809--4817, 2015.
	
	\bibitem{han2019transition}
	Y.~C. Han, Y.~C. Hu, P.~W. Zhang, and L.~Zhang.
	\newblock Transition pathways between defect patterns in confined nematic
	liquid crystals.
	\newblock {\em Journal of Computational Physics}, 396:1--11, 2019.
	
	\bibitem{milnor1969morse}
	J.~W. Milnor, M.~Spivak, and R.~Wells.
	\newblock {\em Morse theory}, volume~1.
	\newblock Princeton university press Princeton, 1969.
	
	\bibitem{robinson2017molecular}
	M.~Robinson, C.~Luo, P.~E. Farrell, R.~Erban, and A.~Majumdar.
	\newblock From molecular to continuum modelling of bistable liquid crystal
	devices.
	\newblock {\em Liquid Crystals}, 44(14-15):2267--2284, 2017.	
	
	\bibitem{tsakonas2007multistable}
	C.~Tsakonas, A. J.~Davidson, C. V.~Brown, and N. J. ~Mottram. 
	\newblock Multistable alignment states in nematic liquid crystal filled wells. 
	\newblock {\em Applied physics letters}, 90(11), 111913, 2007.
	
	\bibitem{kralj2014order}
	S.~Kralj and A.~Majumdar.
	\newblock Order reconstruction patterns in nematic liquid crystal wells.
	\newblock {\em Proceedings of the Royal Society A: Mathematical, Physical and
		Engineering Sciences}, 470(2169):20140276, 2014.
	
	\bibitem{canevari2017order}
	G.~Canevari, A.~Majumdar, and A.~Spicer.
	\newblock Order reconstruction for nematics on squares and hexagons: A
	{Landau-de Gennes} study.
	\newblock {\em SIAM Journal on Applied Mathematics}, 77(1):267--293, 2017.

	\bibitem{yin2020construction}
	J.~Y. Yin, Y.~W. Wang, J.~Z.~Y. Chen, P.~W. Zhang, and L.~Zhang.
	\newblock Construction of a pathway map on a complicated energy landscape.
	\newblock {\em Physical Review Letters}, 124:090601, 3 2020.
	
	\bibitem{han2019reduced}
	Y.~C. Han, A.~Majumdar, and L.~Zhang.
	\newblock A reduced study for nematic equilibria on two-dimensional polygons.
	\newblock {\em SIAM Journal on Applied Mathematics}, 80(4):1678--1703, 2020.
	
	\bibitem{golovaty2017dimension}
	D.~Golovaty, J.~A. Montero, and P.~Sternberg.
	\newblock Dimension reduction for the {Landau--de Gennes} model on curved
	nematic thin films.
	\newblock {\em Journal of Nonlinear Science}, 27(6):1905--1932, 2017.
	
	\bibitem{bethuel1994ginzburg}
	F.~Bethuel, H.~Brezis, F.~H{\'e}lein, et~al.
	\newblock {\em Ginzburg-Landau Vortices}, volume~13.
	\newblock Springer, 1994.

	\bibitem{wojtowicz1975introduction}
	P.~J. Wojtowicz, P.~Sheng, and E.~B. Priestley.
	\newblock {\em Introduction to liquid crystals}.
	\newblock Springer, 1975.

	\bibitem{canevari_majumdar_wang_harris}
	G.~Canevari, J.~Harris, A.~Majumdar, and Y.~W. Wang.
	\newblock The well order reconstruction solution for three-dimensional wells,
	in the {Landau-de Gennes} theory.
	\newblock {\em International Journal of Nonlinear Mechanics}, 119:103342, 2020.
	
	\bibitem{brodin2010melting}
	A.~Brodin, A.~Nych, U.~Ognysta, B.~Lev, V.~Nazarenko, M.~{\v{S}}karabot, and
	I.~Mu{\v{s}}evi{\v{c}}.
	\newblock Melting of {2D} liquid crystal colloidal structure.
	\newblock {\em Condensed Matter Physics}, 2010.
	
	\bibitem{bisht_epl}
	K.~Bisht, Y.~W. Wang, B.~Varsha, and A.~Majumdar.
	\newblock Tailored morphologies in two-dimensional ferronematic wells.
	\newblock {\em Phys. Rev. E}, 101(022706), 2020.
	
	\bibitem{gupta2005texture}
	G.~Gupta and A.~D. Rey.
	\newblock Texture modeling in carbon--carbon composites based on mesophase
	precursor matrices.
	\newblock {\em Carbon}, 43(7):1400--1406, 2005.
	
	\bibitem{Mu2008Self}
	I.~Musevic and M.~Skarabot.
	\newblock Self-assembly of nematic colloids.
	\newblock {\em Soft Matter}, 4(2):195--199, 2008.
	
	\bibitem{Igor2006Two}
	I.~Musevic, M.~Skarabot, U.~Tkalec, M.~Ravnik, and S.~Zumer.
	\newblock Two-dimensional nematic colloidal crystals self-assembled by
	topological defects.
	\newblock {\em Science}, 313(5789):954--958, 2006.

  	\bibitem{jonsson1998nudged}
	H.~J{\'o}nsson, G.~Mills and K.~W. Jacobsen. 
	\newblock  Nudged elastic band method for finding minimum energy paths of transitions.
	\newblock {\em Classical and Quan-tum Dynamics in Condensed Phase Simulations}, World Scientific, Singapore, p.~385, 1998.

  	\bibitem{weinan2002string}
	W.~E, W.~Ren and E.~Vanden-Eijnden.
  \newblock String method for the study of rare events.
  \newblock {\em Phys. Rev. B} 66(5):052301, 2002.

	\bibitem{weinan2011gentlest}
	W.~E and X.~Zhou  
	\newblock The gentlest ascent dynamics.
	\newblock {\em Nonlinearity} 24(6):1831, 2011.

	\bibitem{henkelman1999dimer}
	G.~Henkelman and H.~J{\'o}nsson.  
	\newblock A dimer method for finding saddle points on high dimensional potential surfaces using only first derivatives.
  \newblock{\em The Journal of chemical physics} 111(15):7010--7022, 1999.

	\bibitem{zhang2012shrinking}
	J.~Zhang and Q.~Du.
	\newblock Shrinking dimer dynamics and its applications to saddle point search.
  	\newblock{\em SIAM Journal on Numerical Analysis} 50(4):1899--1921, 2012.

	\bibitem{zhang2016optimization}
	L.~Zhang, Q.~Du, and Z.~Z. Zheng.
	\newblock Optimization-based shrinking dimer method for finding transition
	states.
	\newblock {\em SIAM Journal on Scientific Computing}, 38(1):A528--A544, 2016.
	
	\bibitem{doye2002saddle}
	J.~P.~K. Doye and D.~J. Wales.
	\newblock Saddle points and dynamics of {L}ennard-{J}ones clusters, solids, and
	supercooled liquids.
	\newblock {\em The Journal of Chemical Physics}, 116(9):3777--3788, 2002.
	
	\bibitem{li2001minimax}
	Y.~X. Li and J.~X. Zhou.
	\newblock A minimax method for finding multiple critical points and its
	applications to semilinear {PDE}s.
	\newblock {\em SIAM Journal on Scientific Computing}, 23(3):840--865, 2001.
	
	\bibitem{farrell2015deflation}
	P.~E. Farrell, {\'{A}}.~Birkisson, and S.~W. Funke.
	\newblock Deflation techniques for finding distinct solutions of nonlinear
	partial differential equations.
	\newblock {\em SIAM Journal on Scientific Computing}, 37(4):A2026--A2045, 2015.
	
	\bibitem{mehta2011finding}
	D.~Mehta.
	\newblock Finding all the stationary points of a potential-energy landscape via
	numerical polynomial-homotopy-continuation method.
	\newblock {\em Physical Review E}, 84:025702, 2011.
	
	\bibitem{hao2014bootstrapping}
	W.~R. Hao, J.~D. Hauenstein, B.~Hu, and A.~J. Sommese.
	\newblock A bootstrapping approach for computing multiple solutions of
	differential equations.
	\newblock {\em Journal of Computational and Applied Mathematics}, 258:181--190,
	2014.	

	\bibitem{yin2019high}
	J.~Y. Yin, L.~Zhang, and P.~W. Zhang.
	\newblock High-index optimization-based shrinking dimer method for finding
	high-index saddle points.
	\newblock {\em SIAM Journal on Scientific Computing}, 41(6):A3576--A3595, 2019.

	
	
	
	\bibitem{yin2020searching}
	J.~Y. Yin, B.~Yu, and L.~Zhang.
	\newblock Searching the solution landscape by generalized high-index saddle
	dynamics.
	\newblock {\em arXiv preprint arXiv:2002.10690}, 2020.
	
	\bibitem{fabero2001explicit}
	J.~C. Fabero, A.~Bautista, and L.~Casas{\'u}s.
	\newblock An explicit finite differences scheme over hexagonal tessellation.
	\newblock {\em Applied Mathematics Letters}, 14(5):593--598, 2001.
	
	\bibitem{hu2016disclination}
	Y.~C. Hu, Y.~Qu, and P.~W. Zhang.
	\newblock On the disclination lines of nematic liquid crystals.
	\newblock {\em Communications in Computational Physics}, 19(2):354--379, 2016.
	
	\bibitem{troger2012nonlinear}
	H.~Troger and A.~Steindl.
	\newblock {\em Nonlinear stability and bifurcation theory: an introduction for
		engineers and applied scientists}.
	\newblock Springer Science \& Business Media, 2012.

	\bibitem{yin2020transition}
	J.~Y. Yin, K.~Jiang, A.~C. Shi, P.~W. Zhang, and L.~Zhang.
	\newblock Transition pathways connecting crystals and quasicrystals.
	\newblock {\em arXiv preprint arXiv:2007.15866}, 2020.	

	\bibitem{luo2012multistability}
	 C.~Luo, A.~Majumdar, and R.~Erban.
	\newblock Multistability in planar liquid crystal wells.
	\newblock {\em Physical Review E}, 85(6):061702, 2012.	

\end{thebibliography}
\end{document}